\documentclass[twocolumn,aps,prb,notitlepage,showpacs,floatfix,unsortedaddress,superscriptaddress,citeautoscript]{revtex4-1}
\usepackage{amsmath,amssymb,amstext,graphicx,bm}
\usepackage{graphicx,color,bm}
\usepackage[english]{babel}
\usepackage{dsfont}
\usepackage{epsfig}
\begin{document}
\title{Interplay between strain and  size quantization in a class of topological insulators based on  inverted-band semiconductors}

\author{Alexander Khaetskii}
\email{khaetskii@gmail.com}
\affiliation{Air Force Research Laboratory, Wright-Patterson AFB, Ohio 45433, USA}
\affiliation{KBR, Beavercreek, Ohio 45431, USA}
\affiliation{Department of Physics and Astronomy, Ohio University, Athens,  OH 45701, USA}

\author{Vitaly Golovach}
\affiliation{Centro de F\'{i}sica de Materiales (CFM-MPC), Centro Mixto CSIC-UPV/EHU,  20018 Donostia-San Sebasti\'{a}n, Spain}
\affiliation{Departamento de Pol\'{i}meros y Materiales Avanzados: F\'{i}sica, Qu\'{i}mica y Tecnolog\'{i}a, Facultad de Qu\'{i}mica, University of the Basque Country UPV/EHU, 20080 Donostia-San Sebasti\'{a}n, Spain}
\affiliation{Donostia International Physics Center (DIPC),  20018 Donostia-San Sebasti\'{a}n, Spain}
\affiliation{IKERBASQUE, Basque Foundation for Science, 48013 Bilbao, Spain}

\author{Arnold Kiefer}
\affiliation{Air Force Research Laboratory, Wright-Patterson AFB, Ohio 45433, USA}

\date{\today}

\begin{abstract}
We consider surface states in semiconductors with inverted-band structures, such as $\alpha$-Sn and HgTe. The main interest is the interplay of the effect of a strain of an arbitrary sign  and that of the sample finite size. 
We consider, in particular, a model system comprised of a gapless semiconductor (e.g. HgTe or $\alpha$-Sn) of finite-width  sandwiched between layers of a regular-band semiconductor (e.g. CdTe or InSb). 
We clarify the origin of various transitions that happen at a given strain with the change of the sample thickness, in particular the transition between the Dirac semimetal and quasi-3D (quantized) topological insulator. Our conclusion  opposes those reached recently by the majority of researchers. We show that near the transition point the surface state cannot be treated as a truly topological one since  
the parameters of the problem are such that an appreciable overlap of the surface states' wave functions located at opposite boundaries occur. As a result, a spin-conserving, elastic impurity scattering between the states located at opposite boundaries will induce substantial backscattering and destroy the robustness of the surface state . 
 For the k-p Kane model we derive hard-wall boundary conditions  in the case  when the regular-band materials form high barriers for the carriers of the inner inverted-band semiconductor (for example, CdTe/HgTe/CdTe and CdTe/$\alpha$-Sn/CdTe cases). We show that in this case the boundary conditions have universal and simple form and allow investigation of the realistic case of finite mass of the heavy-hole band, and comparison of the results obtained within the Kane and Luttinger models. In particular, a new type  of surface states (wing states) developes with application of strain in the Kane model and is  absent in the Luttinger model. 
\end{abstract}

\maketitle

 \section{Introduction}      
\label{intr}
Describing the electronic surface states of gapless semiconductors has long been an interesting
problem because of the
 rich physics at their interfaces. 
Gapless semiconductors, like HgTe (or $\alpha$-Sn),
have inverted band structures where the s-like $\Gamma_6 $ ($\Gamma_{7-} $) and p-like $\Gamma_8$ ($\Gamma_{8+}$) bands switch order in energy due
primarily to strong spin-orbit coupling. \cite{Groves,Bloom,Gelmont} 
Crystalline symmetry enforces a four-fold degeneracy at
the  $\Gamma_8$-point, effectively closing the bandgap. Because of the band inversion, interfaces with vacuum or
with direct band-gap semiconductors having ordinary band ordering can produce new types of interface
(surface) states, namely Dyakonov-Khaetskii (DK) states  \cite{Dyak} and Volkov-Pankratov (VP) states \cite{Volkov,Cade,Suris,Volkov1}. The DK states lie above the  $\Gamma_8$ heavy-hole band in HgTe and $\alpha$-Sn materials, whereas the VP branch is located below the  heavy-hole band within the $\Gamma_8- \Gamma_6$ gap.  These two types of states always coexist since they are just two branches of the same topological Dirac state repelled from each other by the coupling to the heavy-hole band.  Moreover, these surface states have nondegenerate spin-texture with a strong spin-momentum locking, making them interesting for
spintronics and optoelectronics applications. \cite{Ando} Although these surface states were predicted in the 1980s,
it has been only recently  that the topological nature of the states was recognized. \cite{Hasan,Qi,Bernevig,Bernevig1,Bernevig2,Kane,Dai,Khaetskii, Pankratov}
 Experimentally, these interface states can be revealed by angle- and spin-resolved photoemission
spectroscopy \cite{Xia,Hsieh}  and magneto-transport measurements. \cite{Brune,Kozlov,Checkelsky,Analytis,Ren,Xia1,Pan} 
\par
The energy dispersion of interface states becomes more complex for many practical samples of a thin topological insulator deposited onto a topologically trivial semiconductor or insulator.  Under these circumstances, both finite-size effects and epitaxial strain  due to a lattice-mismatch between a film of topological material and substrate  can significantly modify the electronic band structure, causing transitions between different topological states.    Because future devices employing topological materials will likely be made using thin films, the interplay of the finite-size effects and strain within a topological insulator is an important and interesting problem.  
\par
In the current literature, there is a broad range of opinions about the origin and exact conditions for topological transitions of thin-film TIs \cite{Ohtsubo,Coster,Anh,Vail}.  
Many researchers who primarily use numerical methods do not provide insight into the physical origin of these transitions,
 because they lack a clear physical model of the  bulk-like and surface-like subbands  involved
 in these transitions. In sections \ref{VP-Gap}-\ref{discus}, we discuss our classification of states which is different compared to those made in Refs. \onlinecite{Ohtsubo,Coster,Anh,Vail}. This allows one to describe correctly the physics of the transitions in question. 
\par  
The main goal of this work is to understand the physics of the series of transitions that happen at a given strain value with changing sample width. In particular, we have considered analytically 
the transition from the Dirac semimetal to the 3D quantized topological insulator that occurs in the presence of a biaxially compressive in-plane strain. In the limit of large width, this strain leads to an overlap of the light- and heavy-hole bands (DSM regime).
We show that the DK states are responsible for  this transition.
Using the exactly solvable  Luttinger  model,  we identify reliably all the bulk-like and surface-like subbands involved in the transition and make concrete predictions about the robustness of the topological states.  
\par
\begin{figure}[!ht]
\vspace{0pt}\includegraphics[width=1.4 \columnwidth]{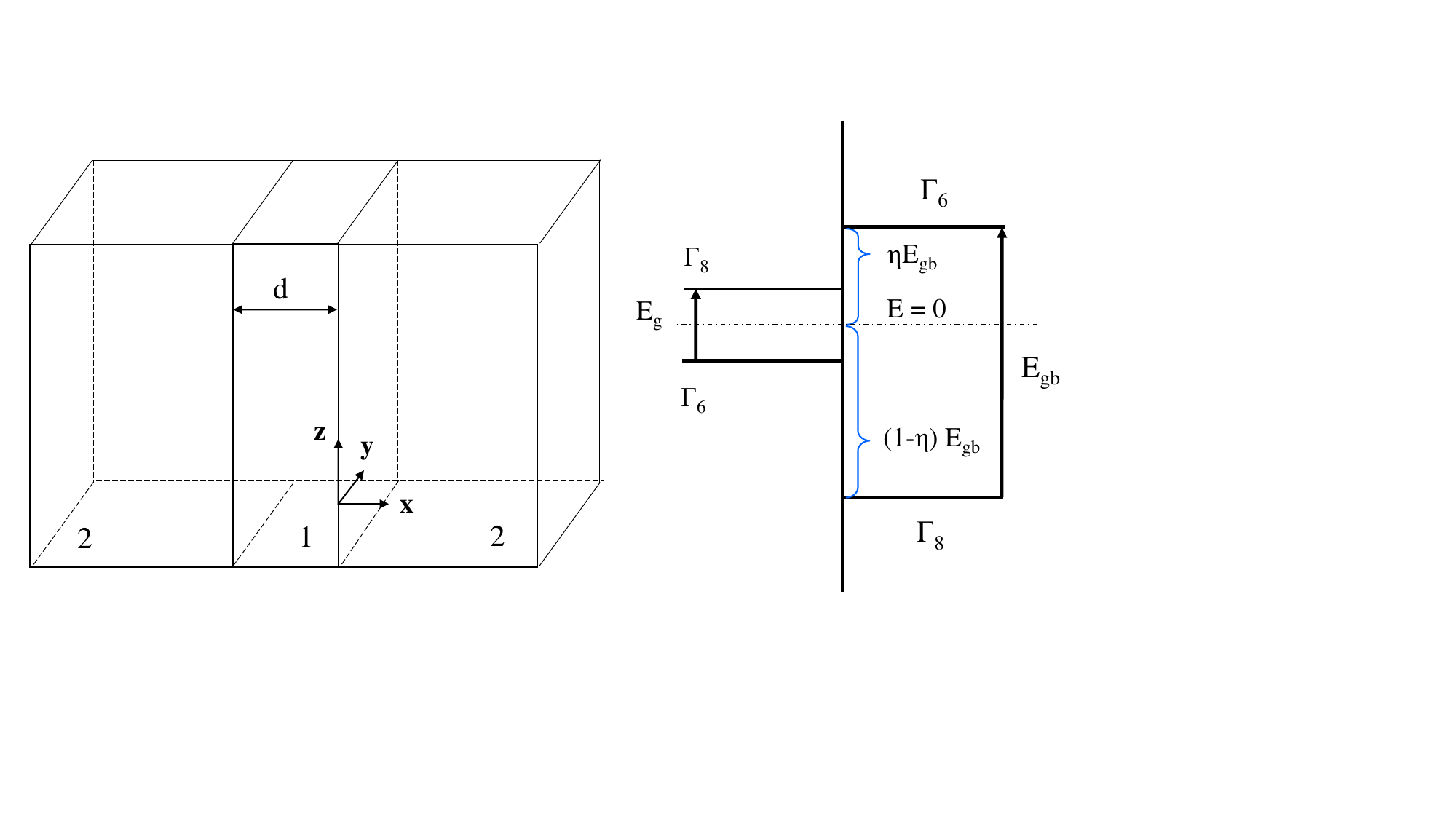}
\caption{\label{Kiefer1}
 Left: Configuration of the structure under consideration.  A gapless semiconductor (region 1) of thickness $d$ is bounded on parallel sides by a regular semiconductor (region 2).  A Cartesian coordinate system is placed such that the $yz$-plane is within one interface where $x = 0$. Right: Idealized energy band diagram of the structure. 
The parameter $\eta$ describes the asymmetry of the offsets in the conduction and valence bands. 
}
\end{figure}

We consider here a model system comprised of a gapless semiconductor with an inverted band structure (e.g. HgTe or $\alpha$-Sn) of finite-width $d$ sandwiched between layers of a semiconductor with a regular band order (e.g. CdTe or InSb), see Fig. \ref{Kiefer1}.
 In section \ref{Problem}, we present the Hamiltonian of the Kane model which includes also the contributions from the far bands; thus the heavy-hole (HH) band acquires a finite mass, which allows one to describe the physics applicable for realistic situations. 
 In section \ref{Wing}, we consider this model with strain to describe the interface between the inverted-band and direct-band  semiconductors focusing on the energies of the DK surface states \cite{Dyak,Khaetskii}. 
In section \ref{DP_loc}, we derive an exact formula for the location of the Dirac point of the VP states (in the case of a single interface).  
In section \ref{VP-Gap}, we introduce finite-size effects and show how they split the VP states and discuss how it relates to possible interpretations of experimental data (for example, ARPES).   In section \ref{Interplay}, we describe the interplay between strain and finite-size effects for cases of biaxially compressive and tensile strain.  Finally, in section \ref{discus}, we discuss our results emphasizing their
difference and novelty compared to the ones
obtained by the other research teams.
In Appendix \ref{BCond}, we develop the necessary boundary conditions (BCs) for a proper solution under finite thickness.  In the case  when the direct-band materials form high barriers for the carriers of the inner inverted-band semiconductor, the BCs  have an universal and simple form,  thereby offering simple analytical treatment and more physical insight than achieved previously.  
\par
  We stress that our main results  address the physics of surface states in  the quasi-3D quantum wells (QWs), 
which differs from the work by Zhou et al. in Ref. \onlinecite{Zhou} where the edge states in the 2D QWs were studied.
The transition from the 2D normal insulator state to the 2D topological insulator state (2D NI-2D TI) occurs at $\simeq $5.8 nm for the CdTe/$\alpha$-Sn/CdTe structure, whereas the DSM to quasi-3D TI transition addressed in this work occurs at $\simeq$25 nm in thickness.
 The first case is discussed in Sec. \ref{VP-Gap} and the second one is considered in Sec. \ref{negative}. The physics corresponding to these cases is totally different in that they involve the coupling of states in different dimensions. The work by Zhou et al. describes the coupling between the 1D edge channels across the width of a 2D sample near the thickness at which the transition 2D NI-2D TI just happened. It is important to mention that for such small thicknesses, strain does not typically play a role.
  The physics we consider in Section \ref{Interplay} in relation to, for example, the DSM-quasi 3D TI transition is necessarily determined by the strain value, and happens at a significantly greater thickness.  Therefore, we consider a quasi-3D sample and study the coupling between the 2D topological surface states.

 \section{Material Configuration and Hamiltonian}    
 \label{Problem}
In this paper we consider several cases related to surface states of a semi-infinite sample or quantized states when one deals with a thin film or QW. In all cases we have a contact of an inverted-band semiconductor (of the $\alpha$-Sn type) with a direct-band semiconductor (like CdTe, for example) or vacuum.  Actual calculations for a finite size sample will be done for the case of symmetric structure of the type  CdTe/$\alpha$-Sn/CdTe. 
\par
The band structures of both the direct and inverted-band semiconductors are described by the $6\times 6$ Kane model which takes into account three bands: electrons (e)  with s-like symmetry, light holes (lh) with p-type symmetry, and heavy holes (hh) also with p-type symmetry.  Using the time-reversal symmetry of the problem and choosing the proper coordinate system reduces the problem to the $3\times 3$ one \cite{Subashiev,Khaetskii}. We choose the angular momentum quantization axis z in the plane of the interface and the direction of the carrier motion along y; x denotes the direction normal to the interface. With this coordinate system, the two groups of states  (e 1/2, lh -1/2, hh 3/2) and (e -1/2, lh 1/2, hh -3/2) do not mix. The Hamiltonian matrix for the first group of states, see Refs. (\onlinecite{Subashiev},\onlinecite{Bir}),
after the use of the unitary transformation 
\begin{eqnarray}
\hat{U}&=&
\begin{bmatrix}
1 & 0 & 0 \\
0 & 1/2&   -\sqrt{3}/2\\
0&   \sqrt{3}/2 & 1/2
\end{bmatrix}
\label{transform}
\end{eqnarray}
acquires the following form: 
\widetext
\begin{eqnarray}
\hat{H}&=&
\begin{bmatrix}
\epsilon_{c} & \frac{ (2\hat{k}_x+ik_y)P}{\sqrt{6}} & iPk_y/\sqrt{2} \\
 \frac{(2\hat{k}_x-ik_y)P}{\sqrt{6}}  & \epsilon_{v} +\Delta/2-\hat{k}_x^2(\gamma+2\tilde\gamma)-k_y^2(\gamma-\tilde\gamma)&
-i\sqrt{3}\tilde\gamma k_y (2\hat{k}_x-ik_y)\\
-iPk_y/\sqrt{2} & i\sqrt{3}\tilde\gamma k_y (2\hat{k}_x+ik_y)  &\epsilon_{v} -\Delta/2 -\hat{k}_x^2(\gamma-2\tilde\gamma)-k_y^2(\gamma+\tilde\gamma)
\end{bmatrix}
\label{Hamilt}
\end{eqnarray}
\endwidetext
The Hamiltonian matrix that refers to the second group of states with the opposite sign of the angular momentum projection on the z-axis can be obtained from Eq.(\ref{Hamilt}) by replacing $k_y$ by $-k_y$. 
In Eq.(\ref{Hamilt}) $P$ is the Kane matrix element which we consider coordinate-independent;  $\gamma, \tilde\gamma$ are second-order valence band Luttinger parameters, which have different values in a gapless semiconductor 1 and in a direct semiconductor 2; and  $\epsilon_{c1,2}, \epsilon_{v1,2}$ are conduction and valence band edge energies (in the absence of strain) in semiconductors 1 and 2, respectively. Strain is applied to the inverted-band semiconductor 1, in practice through an epitaxial constraint.  $\Delta \equiv \Delta_1$ is the strain-induced gap energy between the  light and heavy hole bands and can have either sign, $\Delta_2 \equiv 0$. 

A solution of the equation $(\hat{H}-\epsilon)\Psi=0$ in the bulk case (translation invariance) leads to the equation:
\begin{equation}
H_h(\epsilon)[(\epsilon_c-\epsilon)H_l(\epsilon) -\frac{2}{3}P^2k^2]=\frac{P^2k_y^2\Delta}{2}+3\tilde\gamma k_y^2\Delta (\epsilon_c-\epsilon)
\label{bulk}
\end{equation}
Here $H_l(\epsilon)=\epsilon_v+\Delta/2 -(\gamma+2\tilde\gamma)(k_x^2+k_y^2)-\epsilon$, and $H_h(\epsilon)=\epsilon_v-\Delta/2 -(\gamma-2\tilde\gamma)(k_x^2+k_y^2)-\epsilon$. Throughout the paper we assume the quantities $\gamma_l=(\gamma+2\tilde\gamma)$ and $\gamma_h=(\gamma-2\tilde\gamma)$ in both materials to be positive. 
Equation (\ref{bulk}) determines the bulk spectra of three types of particles: electrons, light holes, and heavy  holes. At zero strain ($\Delta = 0$), when heavy holes are decoupled,  the spectra have the usual, relatively simple form that is determined by equating  two factors on the left hand side of Eq. (\ref{bulk}) to zero. Note that at $k_y=0$, when a motion of particles is perpendicular to the interface, the heavy holes are also decoupled from the light particles (i.e., electrons and light holes). 
\par
In the end of this section let us make the following important remark. 
As it follows from Eq. (\ref{bulk}),  there are only two solutions for $k_x^2$ of the characteristic equation, i.e. a solution for the heavy holes and a solution for light particles (which are the mixture of electrons and light holes). Therefore,  only two independent coefficients exist in the expression for a wave function that is a solution of the Hamiltonian Eq. (\ref{Hamilt}). 
On the other hand, a spinor $\hat \Psi$ has three components. That is why the open boundary condition $\hat \Psi =0$ for the case of high barriers is not valid.  The correct hard-wall boundary condition is derived in Appendix \ref{BCond}. 

\section{Wing states in the Kane model with strain}
\label{Wing}
One of the goals of this work is an attempt to clarify existing ARPES data. In this respect, 
the surface state spectrum near the $\Gamma_8$ energy  for the realistic case of the finite heavy-hole mass is interesting because it allows to exclude or confirm the possibility of forming the twinning of Dirac cones,\cite{Seixas,Balatsky} which we set out to do here for our particular system. 
\par
Two types of surface branches are found within the Kane model. The parabolic branch (called DK1 state) \cite{Dyak,Khaetskii} starts from the HH band and ends at the light-hole (LH) band. The Volkov-Pankratov states lie within the $\Gamma_8-\Gamma_6$ gap \cite{Volkov}. (These two types of states always coexist since they are just two branches of the original topological Dirac state repelled from each other by the repulsive interaction with the heavy-hole band \cite{Khaetskii}).
The  DK1 branch in the presence of strain was studied in Ref. \onlinecite{Khaetskii} for the idealized case of a flat heavy-hole band. Here we investigate this problem at $\Delta>0$ (biaxial, tensile in-plane strain) and $\eta=1/2$  (symmetric band offsets).
The parameter $\eta$ ($0<\eta <1$) describes the asymmetry of the offsets in the conduction and valence bands, see Fig. \ref{Kiefer1}.  We assume 
a small ratio $\beta$ of the light and heavy hole masses given by $\beta=3\gamma_h E_g/2P^2 \ll 1$, $\gamma_h=(\gamma-2\tilde\gamma)$ . We will use the perturbation approach with respect to $\beta \ll 1$. 
\par
We consider the case of a single interface  with a high-energy barrier between the inverted-band and direct band semiconductors.
In the lowest order with respect to small $\beta$ (or equivalently $\gamma_h$)
the spinors in the inverted-band region for the light and heavy particles have the form 
\begin{eqnarray}
\chi_l=\left( 
\begin{array}{cc}
\frac{-i\sqrt{2}B}{Pk_y}& \\
\frac{(-k_y+2\kappa_l)B}{\sqrt{3}k_y (\Delta-B)}& \\
1
\end{array}
\right), \,\,\, \chi_h=\left( 
\begin{array}{cc}
0 & \\
-\frac{\sqrt{3\gamma_h}k_y}{2 \sqrt{B}}& \\
1
\end{array}
\right)
\label{spinors}
\end{eqnarray}
Using the boundary conditions in Eq. (\ref{BCs}), one obtains  the folowing equation  for the surface states spectra
\begin{eqnarray}
\begin{array}{ll}
B^{3/2}[\frac{\sqrt{2}}{P}-\frac{(-k_y+2\kappa_l)}{\sqrt{3}(\Delta-B)}]=\frac{\sqrt{3\gamma_h}}{2}k_y^2,  & \\
\kappa_l=\sqrt{k_y^2+\frac{3}{2}[\frac{(\Delta-B)(\epsilon-\epsilon_c)}{P^2}-\frac{\Delta k_y^2}{2B}]}.
\end{array}
\label{Kane}
\end{eqnarray}
The quantity $B$ in Eqs. (\ref{spinors},\ref{Kane}) is defined as $B=\epsilon+\Delta/2 -E_g/2$. The equation for the time-reversed branch of the states is obtained by replacing $k_y$ by $-k_y$ in Eq. (\ref{Kane}).  
\par
\begin{figure}[!ht]
\vspace{0pt}\includegraphics[width=1.2 \columnwidth]{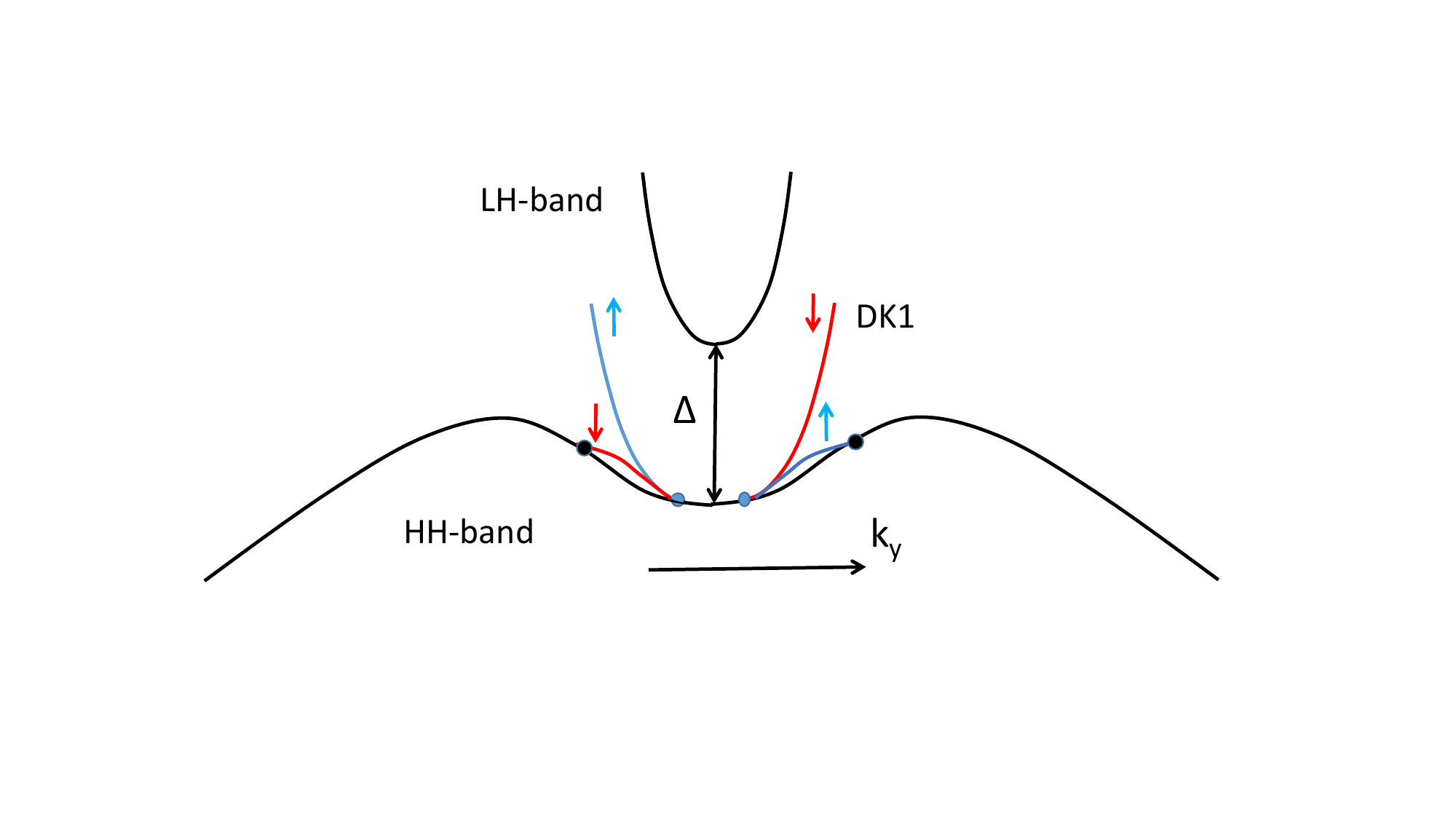}
\caption{\label{Kane_mass}
Surface states around the $\Gamma_8$ point for a semi-infinite sample within the Kane model with a finite HH mass and positive strain, $\Delta >0$. Blue dots correspond to the cut-off $\tilde{k}_y$ points where the DK1 surface branch splits off the HH band, see Eq. (\ref{answer}). Black dots correspond to the $k_{\Delta}$ points, where the wing states terminate. Surface branches are nondegenerate. The spin orientations are shown by the red and blue arrows.}
\end{figure}

For $\beta=0$, using the formulas from Ref. \onlinecite{Khaetskii} we find for the exact spectra of the surface branches 
DK1 which lie above the bottom of the HH band:
\widetext
\begin{eqnarray}
\epsilon_{DK1,\downarrow}(\beta=0)= \frac{(\epsilon_{HH}+\epsilon_{DP})}{2}+\frac{Pk_y}{2\sqrt{6}}+\frac{1}{2}\sqrt{\left (\epsilon_{HH}-\epsilon_{DP}-Pk_y/\sqrt{6}\right )^2+P^2k_y^2}  \nonumber \\
\epsilon_{DK1,\uparrow}(\beta=0)= \frac{(\epsilon_{HH}+\epsilon_{DP})}{2}-\frac{Pk_y}{2\sqrt{6}}+\frac{1}{2}\sqrt{\left (\epsilon_{HH}-\epsilon_{DP}+Pk_y/\sqrt{6}\right )^2+P^2k_y^2}, 
\label{DK1exact}
\end{eqnarray}
where $\epsilon_{HH}=E_g/2-\Delta/2$ is the energy of the bottom of the HH band, and $\epsilon_{DP}=\Delta/4$ is the location of the Dirac point of the Volkov-Pankratov  states (at $k_y=0$).  Note that in contrast to the $\Delta=0$ case, for a finite positive strain there are two DK1 branches corresponding to the opposite spins for a given value of $k_y$. The one corresponding to the spin-up state (for $k_y>0$) exists only up to the critical value of $k_y=k_{\Delta}=\sqrt{6}\Delta/P$ (assuming that $\Delta \ll E_g$), where it merges with the heavy-hole band, see Fig. \ref{Kane_mass} and the discussion below. This branch we call the wing state. 
\par
For completeness we also  present here (at $\beta=0$ ) the exact spectra of two  
VP surface branches that lie within the $\Gamma_8-\Gamma_6$ energy gap (and which we derived using the formulas from Ref. \onlinecite{Khaetskii}):
\begin{eqnarray}
\epsilon_{\downarrow}=  
\frac{(\epsilon_{HH}+\epsilon_{DP})}{2}+\frac{Pk_y}{2\sqrt{6}}-\frac{1}{2}\sqrt{\left (\epsilon_{HH}-\epsilon_{DP}-Pk_y/\sqrt{6}\right )^2+P^2k_y^2}
\nonumber \\
\epsilon_{\uparrow}= 
\frac{(\epsilon_{HH}+\epsilon_{DP})}{2}-\frac{Pk_y}{2\sqrt{6}}-\frac{1}{2}\sqrt{\left (\epsilon_{HH}-\epsilon_{DP}+Pk_y/\sqrt{6}\right )^2+P^2k_y^2}
\nonumber \\
\label{Dirac}
\end{eqnarray}
\endwidetext
Note that Eq. (\ref{Dirac}) is the generalization for the case of $\Delta \neq 0$ of the equations of Ref. \onlinecite{Volkov1}.  
Now we solve Eq. (\ref{Kane}), looking for $B$ by the perturbation theory in small $\beta$; for that we use the expression Eq.(\ref{DK1exact}) while calculating the factor $B^{3/2}$. 
We consider the case $\Delta, Pk_y \ll E_g$. Then for quantity $B_0\equiv B|_{\beta=0}$ it is enough to use a simple expression
$B_0=P^2k_y^2/2E_g$. Using the notation $M=P^2k_y^2/2B$, and assuming $B\ll \Delta$ (which actually means that $Pk_y \ll \sqrt{E_g \Delta}$), we obtain the desired spectrum of the DK1 surface branch 
\begin{eqnarray}
M&=&E_g-\Delta/2+\Delta (1-\tilde k_y/|k_y|)(\tilde k_y/|k_y|-1-2 k_y/k_{\Delta }), \nonumber \\
\epsilon_{DK1,\downarrow}&=&\frac{E_g}{2}-\frac{\Delta}{2} + \frac{P^2k_y^2}{2M} 
\label{answer}
\end{eqnarray}
Here $\tilde k_y=\sqrt{2\beta} E_g/P, \, \, k_{\Delta}=\sqrt{6}\Delta/P$. 
We will consider the case of small $\beta \ll (\Delta/E_g)^2 \ll 1$, then two characteristic momenta $\tilde k_y$ and $k_{\Delta}$ relate as $\tilde k_y \ll k_{\Delta}$. The first one, $\tilde k_y$, is the cut-off value at which the spin-down DK1 branch starts splitting off the HH band, see Fig. \ref{Kane_mass}. (At $k_y<\tilde k_y$ the quantity $\kappa_l$, Eq. \ref{Kane},  is not  real). 
Recall that for $\beta=0$ case the starting point was zero $k_y$ value \cite{Khaetskii}.  At large values of $k_y$ this branch is described by the first line of Eq. (\ref{DK1exact}). 

\par
The second (wing) branch of surface states in this energy interval near the $\Gamma_8$ point corresponds to the spin-up state and is obtained by changing the sign of $k_y$ in Eq. (\ref{answer}). This branch (at $k_y >0$ ) also starts near $\tilde k_y$ (with slight shift from this point determined by small quantity $\tilde k_y /k_{\Delta}\ll 1$), has nonmonotonic behavior,  and merges again with the heavy-hole band at the point $k_{\Delta}$, Fig. \ref{Kane_mass}.  This wing branch exists within the Kane model already in the case of $\beta=0$, see the second line of Eq. (\ref{DK1exact}).  Since the interval of $k_y$-values at which it shows up is narrow and is determined by a small strain value, we did not describe this branch in our previous work, Ref. \onlinecite{Khaetskii}. (Note  that within the Luttinger model, when $E_g \to \infty$, the wing branches  disappear and  Fig. \ref{Kane_mass} reduces to Fig. 6(b) of Ref. \onlinecite{Khaetskii}). 
Thus, based on the results obtained here (Fig. \ref{Kane_mass}), we can exclude the formation of a twin Dirac cone for the system we consider. 
\par
A similar consideration for the case $\Delta=0$ leads to the spectrum
$$
\epsilon_{DK1}=\frac{E_g}{2}+\frac{P^2k_y^2}{2E_g}-\frac{P^2k_y^2 P(\tilde k_y-k_y)}{\sqrt{6}E_g^2}
$$
Here $P\tilde k_y, Pk_y \ll E_g$.
We note that this branch proceeds up to the $k_y=0$ point since the edge of the bulk heavy-hole band in this case is described by the equation $\epsilon_h(k_y)=E_g/2 -\gamma_h k_y^2$ (i.e. goes down with $k_y$ from the very beginning).

\section{Location of the Dirac point of the Volkov-Pankratov states}
\label{DP_loc}
 Existing ARPES data reveal two experimental facts that strongly disagree with the theoretical predictions of many papers.
Some ARPES experiments \cite{Engel,Polaczynski} with $\alpha$- Sn material show that the DP lies in close proximity to the $\Gamma_8$ point. From  Eq. (\ref{DP}) it follows that for the actual parameters of the structures which have the  energy diagram of the type presented in Fig. 1 this fact cannot be explained.  
The other puzzling experimental fact is that the dispersion of the surface states (presumably of the VP types) appears  linear in k-vector and connect s-type band with the light-hole band crossing the fundamental gap without experiencing any repulsive interaction with the bulk HH band. 
The last fact contradicts the theoretical predictions, see, for example, Ref. \onlinecite{Khaetskii}.

\par
For {\it arbitrary} values of the gap energies, $E_g>0, E_{gb}>0$, in the inverted and direct (barrier) materials, and arbitrary offsets in the conduction and valence bands, we obtain the following formula for the location of the Dirac point of the Volkov-Pankratov states \cite{Volkov} 
\begin{equation}
\epsilon_{DP}=\frac{\Delta}{2}\eta + \frac{E_g E_{gb}}{E_g+E_{gb}}(\eta-1/2)
\label{DP}
\end{equation}
A similar result was obtained earlier in Ref. \onlinecite{Cade}. 
The zero of the energy coincides with the middle of the 
$\Gamma_8-\Gamma_6$  gap of the inverted material. 
While deriving Eq. (\ref{DP}) we neglected the $\gamma,\tilde\gamma$ terms in the Hamiltonian and used the continuity of the electron and light-hole components of spinors found at $k_y=0$ in the inverted-band and direct-gap  materials. 
\par
The location of  the DP close to the $\Gamma_8$ point can be obtained from Eq. (\ref{DP}) in the case when $E_{gb} \gg E_g$ and $\eta \to 1$. 
$\eta \to 1$ means that the offset in the conduction band is much larger than in the valence band. 
Then at $\Delta=0$ one obtains $\epsilon_{DP}=E_g/2$, i.e. at the energy of the $\Gamma_8$ point of the inverted material. 
\par
If, however, one considers the case of $\alpha$-Sn/InSb interface \cite{Engel}, then using the values of offsets from
 Ref. \onlinecite{Cardona} and assuming a zero strain, one obtains the location of the DP to be  rather closer to the middle of the $\alpha$-Sn gap, which differs from the interpretations \cite{Engel} of the ARPES measurements. 
It should be mentioned that the situation does not become better even if an interaction with the far bands is taken into account. 
 In Appendix \ref{D1} the correction to Eq. (\ref{DP}) due to the terms proportional to $\gamma$ and $\tilde \gamma$  in the Hamiltonian  Eq. (\ref{Hamilt}) is derived. We have shown  that such correction is negative for the cases 
we consider in this work (HgTe/CdTe and $\alpha $-Sn/InSb), i.e. a shift of the DP is downward in energy with respect to the location given by Eq. (\ref{DP}) and cannot explain the experimental data. 
\par
It is not clear if Eq. (\ref{DP}) has an analogue in the case the boundary is with vacuum. 
Interestingly, the authors of Ref. \onlinecite{C. Liu} claimed that a strong reconstruction of the surface \cite{Z. Lu}  occurs  for their bulk HgTe when bounded by vacuum which leads to the opening of a large gap between the light and heavy-hole bands. Mathematically it is equivalent to the presence of a big positive strain $\Delta$.  Such a big positive strain can indeed push the DP towards the top of the heavy-hole band. 
\par
We suggest both experimental facts mentioned in the beginning of this section cannot be explained  within the idealized theoretical model presented by Fig. 1.  Recall that 
for the typical range of ARPES excitation energies, the escape depth of electrons  is about 1 nm near the interface. Here we refer to the so-called 'universal curve" for all electron spectroscopies including ARPES. \cite{Seah_Dench}
Consequently, information about the surface states is collected only within $\sim$ 1 nm near the vacuum surface.  If a bending of the energy bands near the interface in the inverted-band material (HgTe or $\alpha$-Sn) occurs in such a way that it creates a potential barrier for the heavy holes, then they just cannot reach the interface layer which is probed by ARPES. Perhaps only the heavy holes that travel exactly perpendicular to the surface can tunnel into the interface layer, but they are decoupled from the light particles anyway \cite{Khaetskii}.  Under such conditions the ARPES pattern will look like in Fig. 1 of Ref. \onlinecite{Khaetskii}. 
Moreover, the picture of the band bending described above means that  the energy of the $\Gamma_8$ point in the inverted-band material  is shifted downward, and the DP of the VP states may be close to the $\Gamma_8$ point.

\section{Gap in the  Volkov-Pankratov states}
\label{VP-Gap}
Above we have found the location of the Dirac point of the 
Volkov-Pankratov states as a function of the bands' parameters of the neighbouring materials  in the case of a single interface.           A gap is expected to open in the VP states in the case when a film width is finite because of a hybridization of the states belonging to the opposite boundaries.  
This quantity is also important for the interpretation of the ARPES data. 
  We now consider the effects of a sample finite size on the gap value and spectrum of the VP states.  We will be interested in the VP states spectra for small values of $k_y$,  at least $k_y \ll E_g/P$, thus we can neglect the interaction with the heavy holes. For simplicity we consider  also the zero strain case.  Under these conditions, from 
 Eq. (2) of Ref. \onlinecite{Khaetskii} we obtain the following two-band model Hamiltonian
\begin{eqnarray}
\hat{H}&=&
\begin{bmatrix}
\epsilon_{c1,2} & \frac{ (2\hat{k}_x+ik_y)P}{\sqrt{6}} \\
\frac{ (2\hat{k}_x-ik_y)P}{\sqrt{6}} &\epsilon_{v1,2}
\end{bmatrix}
\label{2bands}
\end{eqnarray}
For the film of the width $d$ with the high barriers using the boundary conditions Eq. (\ref{BCs})
one obtains the following equation for the VP surface states   spectrum
\begin{eqnarray}
\tanh(\alpha \xi)=\xi   \frac{E_g/s}{E_g+2\epsilon \rho}; \,\, \alpha=\frac{\sqrt{3}E_g d}{2\sqrt{2}P}, \,\,s=\frac{1}{2\sqrt{\eta (1-\eta)}}, \nonumber \\
 \rho=1-2\eta, \,\, \xi(\epsilon)=\sqrt{1-4(\epsilon/E_g)^2+2P^2k_y^2/3E_g^2}  \nonumber \\
\label{gap}
\end{eqnarray}
In the limit of infinite width $\tanh(\alpha \xi)=1$, and the solution of the quadratic Eq. (\ref{gap}) reads 
\begin{equation}
\epsilon_{\infty}=-\frac{\rho E_g}{2} \pm \frac{Pk_y}{\sqrt{6}s}; \,\, \xi_{\infty}=\frac{1}{s} \pm \frac{2\rho Pk_y}{\sqrt{6}E_g}
\label{VP_infty}
\end{equation}
It is interesting that the velocities of the VP branches depend on the parameter $\eta$ which describes the asymmetry of the offsets in the conduction and valence bands. 
In the limit of large width $(\alpha /s \gg 1)$, the solution is $\xi \to \xi_{\infty}$ and the spectrum is 
\begin{eqnarray}
\epsilon=E_g (\eta-1/2)\pm \frac{1}{2}\sqrt{\Delta_{\pm}^2+ 2P^2k_y^2/3s^2}, \nonumber \\
\Delta_{\pm}=\frac{2E_g}{s^2} e^{-\alpha(1/s\pm 2\rho Pk_y/\sqrt{6}E_g)} 
\label{gap1}
\end{eqnarray}
Thus in this limit the gap value is exponentially small with respect to the film width. For the derivation of Eq. (\ref{gap1}) see  Appendix \ref{A1}.
 Eq. (\ref{gap1})  shows that reducing $d$ leads to the opening of a gap in these states with the upper branch increasing in energy and the lower branch decreasing, see also Fig. 3 of the  Ref. \onlinecite{Subashiev}. One can say that the location of the  central point of the VP branches stays at rest within the first-order approximation, but a gap opens between the two Dirac cones, see Appendix \ref{A1}.  
\par
It is important to realize what happens near the critical width $d_1$ when the upper branch of the Volkov-Pankratov states ($S1$ branch in Fig. 4 of Ref. \onlinecite{Subashiev}) is inverted with the $hh1$ sub-band, which is the first quantized sub-band of the heavy holes (see also Ref. \onlinecite{Bernevig1}).  Near this width the energy spectrum (Fig. 4(b) of Ref. \onlinecite{Subashiev}) will look like a linear-in-$k_y$ Dirac dependence with no gap (see also Ref. \onlinecite{Dietl}). One might erroneously conclude that near this energy one has Volkov-Pankratov states with DP close to the $\Gamma_8$ point. 
To the contrary, the lower branch of the Volkov-Pankratov states (i.e. $S2$ branch, see Figs. (3,4) of Ref. \onlinecite{Subashiev}) lies well below this energy.  Therefore,  the "DP" (i.e. the energy where both $S1$ and $S2$ branches match each other in the limit of a very large thickness) also lies below. Note that at $d > d_1$ (see Fig. 4(b) of Ref. \onlinecite{Subashiev}) the inverted $hh1$ state coincides with the DK1 surface state at $k_y d > \pi $ . For the details see Ref. \onlinecite{Dyak1}.

\begin{figure*}[!ht]
\vspace{0pt}\includegraphics[height=15em]{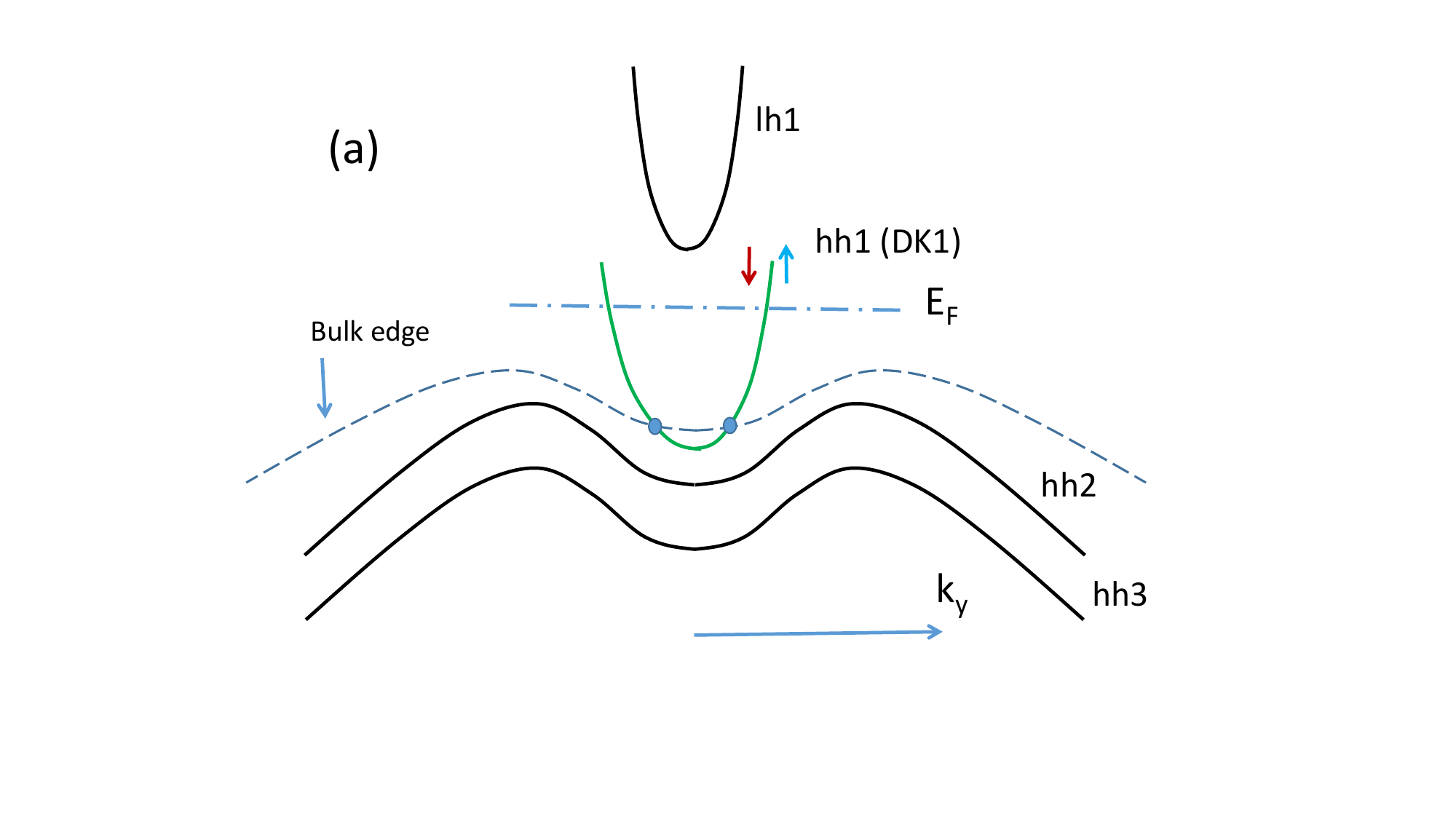}
\hspace{2pt}
\vspace{0pt}\includegraphics[height=15em]{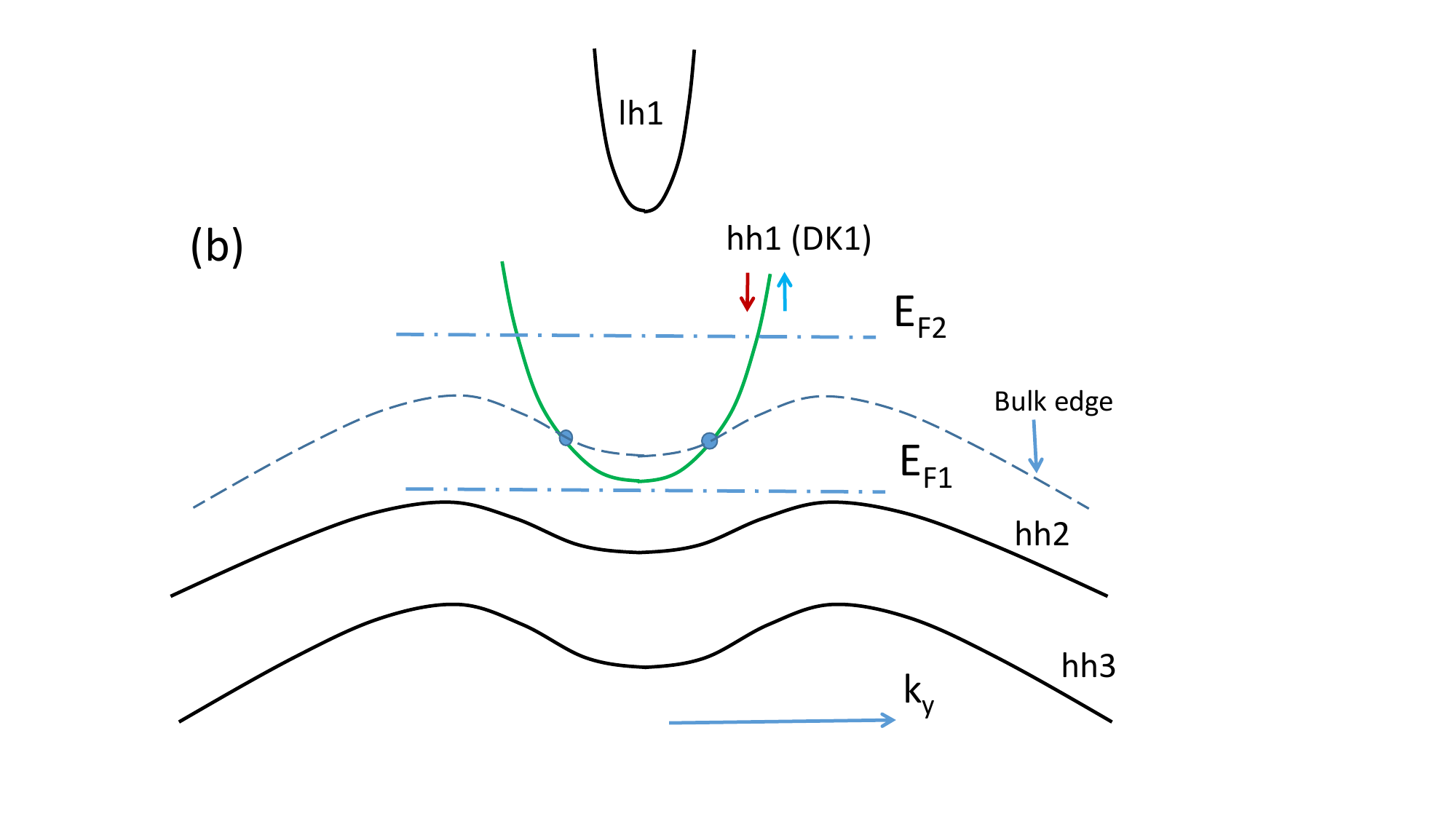}
\caption{\label{size2}
 Bulk-like sub-bands and surface states existing near the $\Gamma_8$ point for a positive strain, $\Delta>0$, and a finite sample width.  The dashed line is the edge of the bulk heavy-hole band in the presence of strain.  Blue dots indicate the $k_y^{\star}$ points where the surface DK1 state transforms into the bulk-like $hh1$ subband, see Eq. (\ref{kstar}). Dashed-dot lines indicate different locations of the Fermi level, see the text.  The surface branch is doubly degenerate. (a)  The case  $d>d_c$.
(b) The case  $d<d_c$.  Width $d_c$ is defined by Eq. (\ref{d_c}). 
 }
\end{figure*}

\section{Interplay between the finite size effects and applied strain}
\label{Interplay}

As we have already mentioned in the Introduction, the most important thing in the description of the transitions related to the competition between the finite size effects and applied strain is the correct identification of the subbands involved in the corresponding transitions. It is not enough just to find the width of the film where some subbands cross each other and claim that it is the point of the transition. One needs also to investigate the dispersions of the corresponding subbands as the function of the in-plane momentum. Depending on the character of these dispersions, whether they are of the normal or an inverted type, the description of the physics is very different. For example, the authors of Refs. (\onlinecite{Coster, Anh,Vail}) did not take this fact into account which led them, in particular, to an incorrect conclusion about the physical meaning of various gaps formed in the system under the size quantization, see below. 
\par
Moreover, none of the papers Refs. (\onlinecite{Ohtsubo,Coster,Anh,Vail}) investigates a robustness of the topological surface states near the transition point, which  is determined by the overlap of the wave functions of the surface states located near the opposite boundaries. This issue is considered for the first time in the current work. We have shown (see discusson below and Fig. \ref{J}) that the parameters of the problem  are such that an appreciable overlap of the surface states' wave functions located at opposite boundaries occurs. 
As a result, an elastic (and spin-conserving) impurity scattering between the states located at opposite boundaries will destroy the robustness of the surface state. 
\par 
In the following, we consider a film of the width $d$ with a center of inversion (identical boundaries).  Subsections A and B below work through the positive and negative cases of strain, respectively.
\par 
The important questions regarding the transitions considered in this section can be answered already  within the Luttinger model, while considering  the characteristic energies near the $\Gamma_8$ point. Moreover, the essential physics can be studied within the model with high barriers, then the BCs reduce to  the zero values for both light-hole and heavy-hole components; see the corresponding discussion in Appendix \ref{BCond}, after Eq. (\ref{applicab}). Everywhere in this section the origin of the energy is at the $\Gamma_8$ point.
We note that all solutions are doubly degenerate with respect to spin because of inversion symmetry. 
As a measure of  topological protection of the surface states at opposite boundaries, one can choose to consider  $1-J$,
where $J$ is the following overlap integral
\begin{eqnarray}
J=\left\langle\psi_{-k_y}|\psi_{k_y}\right\rangle,
\label{defJ}
\end{eqnarray}
where $\left|\psi_{k_y}\right\rangle$ are the states in the quantum well omitting the Bloch exponent $e^{i k_y y}$.
The state $\left|\psi_{-k_y}\right\rangle$ in Eq.~(\ref{defJ}) is chosen from the same block of the Hamiltonian as the state $\left|\psi_{k_y}\right\rangle$
and hence the spinor structure of the two considered states at opposite boundaries is alike \cite{noteintegral}.

\subsection{Positive (biaxially tensile in-plane)  strain, $\Delta >0$}
\label{positive}
We begin with the case of positive strain when for a semi-infinite sample one has a 3D topological insulator regime. We now consider what happens with a reduction of the sample size $d$. 
The  limiting cases of $d\to \infty$ and $\Delta=0$ were studied before, and the pictures for these cases are presented in Figs. 6(b) of Ref. \onlinecite{Khaetskii} and Fig. 2 of Ref. \onlinecite{Dyak1}, respectively.

\par
Let us consider first the limit of large width $d$, when the typical quantization energy of light or heavy holes  is small compared to the strain-induced gap, $\hbar^2 \pi^2/\mu d^2 \ll \Delta, \,\, 1/\mu=1/m_l +1/m_h$. Here $m_l>0$ and $m_h>0$ are bulk masses of light and heavy holes. 
The light  and heavy-hole bulk states form a set of sub-bands, see Fig. \ref{size2}(a).  
A key point is that for any finite width $d$  the quantized states can exist at arbitrary values of in-plane momentum $k_y$, up to zero value. This differs from the semi-infinite case described above where the DK1 surface branch ceases to exist at some critical $k_y$ value (see Fig. \ref{Kane_mass} of the present paper and Fig. 6(b) of  Ref. \onlinecite{Khaetskii}).   
This difference is due to the fact that zero boundary conditions on both boundaries of a {\it finite} sample can be satisfied with the functions of the form $\sin k_x x, \cos k_x x$.
\par
 As a result, the DK1 branch is transformed into  the first  heavy-hole subband $hh1$, see Fig. \ref{size2}. This transformation occurs at the  $k_y=k_y^{\star}$ value which is found from the condition 
$\kappa_l=0$ (the expression for $\kappa_l$  and the exact equation for  $k_y^{\star}$ obtained from this condition  are given in Appendix \ref{A2} by Eq. (\ref{kappa}) and Eq. (\ref{kstar}), respectively). 
At this point the DK1 state crosses the edge of the bulk 
heavy-hole band.  One can also say that at the point $k_y^{\star}$ the  heavy-hole components of the $hh1$ state transform from the form $\sin k_x x, \cos k_x x$ (localized near the center of the film) to the exponents which are localized more near the boundaries (i.e. $hh1$ bulk-like state transforms into the DK1 surface state), see also Ref. \onlinecite{Dyak1}.
\par
 Allowed values  of transverse momenta at $k_y=0$ are $k_x^{(n)}=\pi n/d$ and the corresponding energies of the light and heavy-hole subbands are $E_{\textsl{lhn}}=\hbar^2 \pi^2 n^2/2m_{l} d^2 +\Delta/2$ and $E_{\textsl{hhn}}=- \hbar^2 \pi^2 n^2/2m_{h} d^2 - \Delta/2$, respectively, see Fig. \ref{size2}. 
\par
\begin{figure}[!ht]
\vspace{0pt}\includegraphics[height=14em]{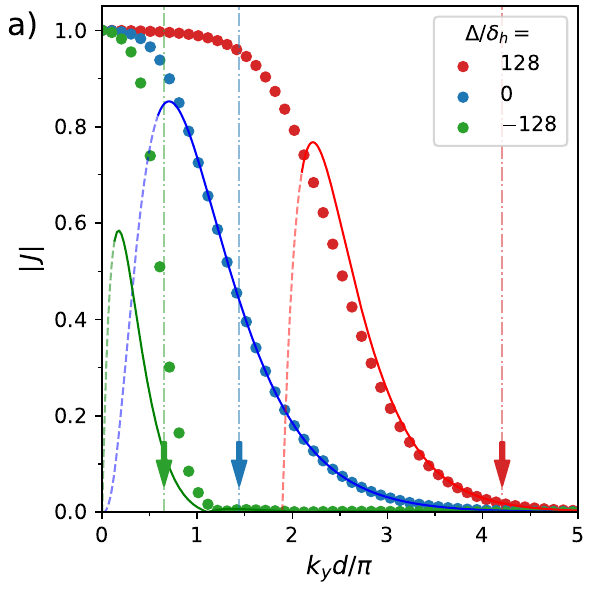}
\vspace{0pt}\includegraphics[height=14em]{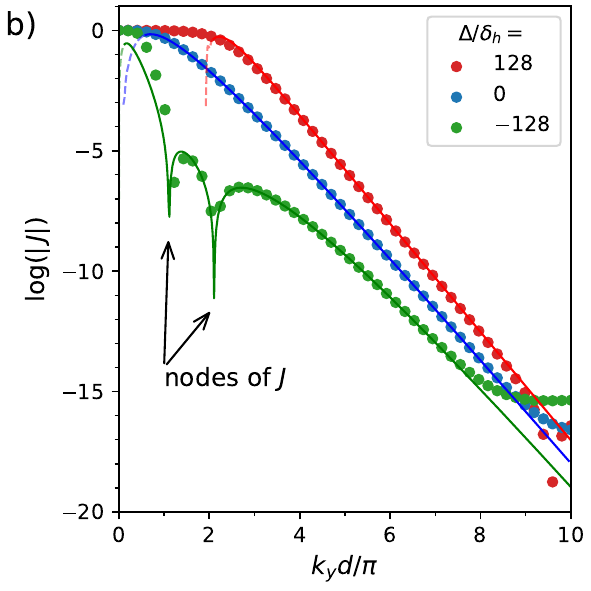}
\caption{\label{J}
 The overlap integral $J$ in Eq.~(\ref{defJ}) calculated for the $hh1$ (DK1) subband in a HgTe quantum well for different values of strain. 
 The strain parameter $\Delta$ is measured in units of the heavy-hole quantization energy $\delta_{\textit{h}}=\hbar^2\pi^2/2 m_{h}d^2$.
 (a) The filled circles show the overlap obtained from a numerical diagonalization of the Luttinger Hamiltonian, whereas the solid lines show the result of 
 the weak-overlap approximation, which is continued by dashed lines in the regime where the approximation breaks down.
 The arrows with dash-dotted vertical lines in (a) show the values $k_y^{\text{max}}$ at which the 
bottom of the bulk-like $lh1$-subband equals to the energy of the surface-like DK1 state.
(b) Same as in (a), but on a logarithmic scale.  Parameters of HgTe: $m_h=0.48\, m_0$ and $m_l=0.03\,m_0$. 
 }
\end{figure}

The parameter which determines the crossover between the regimes with an exponentially small value of $J$ and the value of $J\simeq 1 $ is  $\theta_d=d \sqrt{m_l \Delta}/\pi \hbar$. (Everywhere in this section we assume $\beta=m_l/m_h \ll 1$).
At large width $d$ such that $\theta_d \gg 1$,  Fig. \ref{size2} (a) describes a material which  still can be treated as a 3D topological insulator. 
For that, the Fermi level should lie within the $hh1$  branch above the edge of the bulk heavy-hole band (shown by the dashed line) and below the bottom of the $lh1$ sub-band. There are no bulk states at the Fermi level for this situation, furthermore the overlap $J$ of the surface DK1 states belonging to the opposite boundaries is exponentially small since  $\theta_d \gg 1 $. This case of a big strain corresponds to the red curve of Fig. \ref{J}, where the overlap integral $J$ of two DK1 surface states localized near the opposite surfaces and having opposite $k_y$ vectors is plotted as a function of $k_y$ for HgTe. The maximally allowed value of $k_y$ ($k_y^{\text{max}}$)  corresponding to the energy of  the bottom of the $lh1$ sub-band is shown by the red arrow. We see that for parameters corresponding to the red curve (which can be realized for realistic values of strain only for relatively wide films $d> 70 nm$) one has the lowest value of the overlap integral $J \approx 0.02$. 
\par
 For thinner films the  value of $J$ will be much larger. The red curve will  become closer to the blue one corresponding to zero strain, and $k_y^{\text{max}}$ from being proportional to $\sqrt{\Delta}$ at large $\Delta$ (and large $d$) will change  with decreasing $d$ tending  to $\simeq \pi/d$. For example, for a width $40 nm$ and the strain value corresponding to $\Delta=20 meV$ the parameter $\Delta/\delta_h$ is around 40, and the overlap integral value for a maximally allowed $k_y^{\text{max}}$ is $J\approx 0.25$. (Here $\delta_{\textit{h}}$ is the heavy-hole quantization energy, $\delta_{\textit{h}}=\hbar^2\pi^2/2 m_{h}d^2$).
\par
We note that at zero strain 
while keeping the Fermi level location below the bottom of the $lh1$ sub-band one always has $k_y^{\text{max}} d /\pi \simeq 1$, and since for the DK1 state the length of the localization near the boundary is $\simeq 1/k_y$ \cite{Dyak}, the value of $J (\approx 0.4)$ is automatically of the order of unity.  
\par
Observe that with increasing the positive strain value the overlap integral increases (the red curve versus a blue one) which is quite a  counterintuitive behavior. To understand it we note that 
in the case of exponentially small overlap one has the following analytical expression for $J$ (in the case of $\beta \ll 1$)
$$
J \simeq 2\kappa_l d e^{-\kappa_l d}; \kappa_l=\frac{k_y}{2}[1-\frac{(k_y^{\star})^2}{k_y^2}]; \,  (k_y^{\star})^2=\sqrt{\frac{\beta}{3}} \frac{\Delta}{\tilde \gamma}, k_y>k_y^{\star}.
$$
We have obtained the above expression for the decaying factor of light holes $\kappa_l$  using Eqs. (26, 27)
of Ref. \onlinecite{Khaetskii}. Thus we see that the decaying vector of light holes decreases with strain, in opposition to what might naively expect.

\par
 At some critical width $d_c$ the local maxima at $k_y \neq 0$ of the $hh2$ sub-band will become lower than the bottom of the $hh1$ sub-band. It happens when 
\begin{equation}
3\hbar^2 \pi^2/2m_h d_c^2    \simeq \Delta/4,  
\label{d_c}
\end{equation}
assuming that $\beta \ll 1$.
At smaller values of $d$ it is possible to tune the Fermi level $E_{F1}$  between $hh1$ and $hh2$ sub-bands. Then at $T=0 K$ there are no states at the Fermi level, and a sample constitutes a normal insulator, see Fig. \ref{size2}(b). 
(To avoid confusion, we do not consider here the surface states which presumably exist on the side edges of the sample, and only mean the excitations which can be created in the "bulk" of this quasi 3D sample). 
On the other hand, if the Fermi level $E_{F2}$ is located again within  the $hh1$  branch (above the edge of the bulk heavy-hole band and below the bottom of the $lh1$ sub-band), then only surface states are at the Fermi level, however under the condition indicated above their overlap is not exponentially small. Thus the situation shown in Fig. \ref{size2}(b)  corresponds rather to a normal insulator (semiconductor).  
\par
The general conclusion in the case of positive strain  is that for the film widths within the  usual experimental range of $10 nm <d<100 nm$ and for realistic strain values the overlap integral of the DK1 states belonging to the opposite boundaries is not small and spin-conserving backscattering between them is not suppressed. As a result, it will destroy the robustness of the surface state.  

\begin{figure}[!ht]
\vspace{0pt}\includegraphics[width=1.2 \columnwidth]{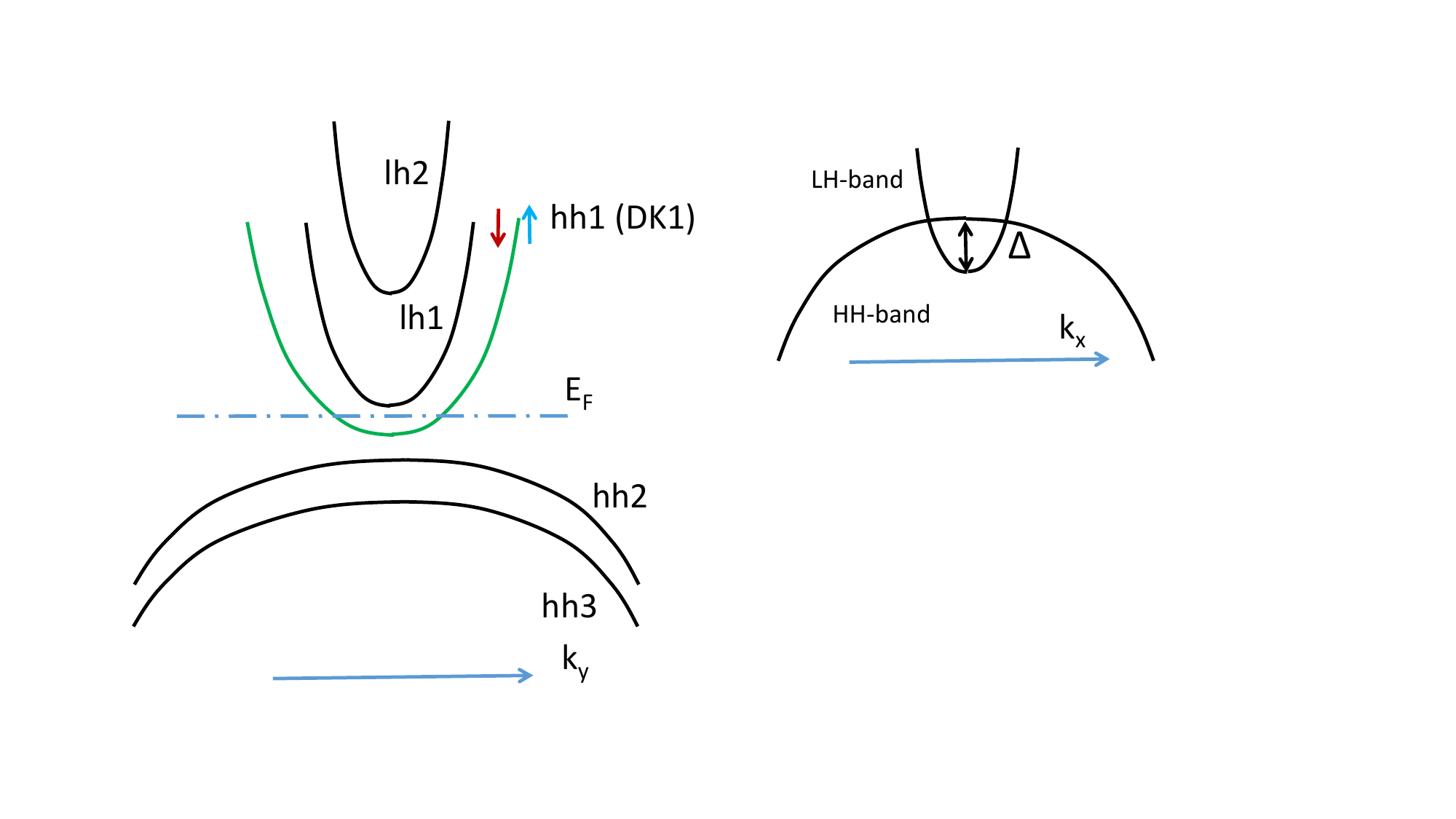}
\caption{\label{size4}
Several sub-bands of bulk-like and surface states shown near $\Gamma_8$ point for a negative strain, $\Delta<0$, and finite sample width $d < d_{|\Delta |} $.  $d_{|\Delta |}$ is defined by Eq. (\ref{width}).  The surface branch is doubly degenerate. The dashed-dot line indicates the location of the Fermi level discussed in the text. 
The inset shows a dispersion of the bulk light and heavy-hole bands as a function of the normal  component $k_x$ of the wave vector.
}
\end{figure}

\subsection{Negative (biaxially compressive in-plane) strain, $\Delta <0$}
\label{negative}
 Let us consider biaxial, compressive  in-plane strain. In accord with the definition we used in Ref. \onlinecite{Khaetskii} it corresponds to the strain-induced energy $\Delta <0$. Then for a large enough sample without size-quantization effects we have a Dirac semi-metal  (DSM) phase. We consider here the transition between the DSM phase and the quasi 3D (quantized) topological insulator phase (as it is called in the literature \cite{Coster,Vail}) through the reduction of the layer width $d$.  We will argue that below some critical width $d_{|\Delta |}$ the DSM transforms into an insulator  which, strictly speaking, cannot be considered  as topological for reasons similar to the ones described in the previous subsection.     We consider the width in the interval 
  \begin{equation}
d_1 \ll d < d_{|\Delta |}, \,\, \frac{\hbar^2 \pi^2}{2\mu d_{|\Delta |}^2} = |\Delta |, \, \, \frac{1}{\mu}=\frac{1}{m_l }+
\frac{1}{m_h},
\label{width}
\end{equation}                                              
where  $d_1$ is the width at which  the upper branch of the Volkov-Pankratov states becomes inverted with the first quantized sub-band of the heavy holes $hh1$, see Ref. \onlinecite{Subashiev}.  
$ d_{|\Delta |}$ is the characteristic sample width at which the strain-induced energy is equal to the sum of the quantization energies of the light and heavy holes.  (In the case of HgTe and for a realistic strain value corresponding to $\Delta=20 meV$ the length $d_{|\Delta |}\approx 26 nm $.)
The left inequality in Eq. (\ref{width}) means that the width is already large enough so that both branches of the Volkov-Pankratov states lie well below the energy of the $\Gamma_8$ point and we can describe the problem within the Luttinger model. 
 The right inequality in Eq. (\ref{width}) means that the width is small enough to “quench” the DSM state (i.e. the energy due to the size quantization is larger than the original overlap of the light and heavy-hole bands due to strain), and the transition to a new state has already happened. 
Our task in this section  is to understand the properties of this state. The physical model presented in recent publications (see, for example, Ref. \onlinecite{Vail}) is unclear, especially  the nature  of "surface" and "bulk" gaps discussed there.  
\par
We argue that 
the physics for this situation was described in Refs. (\onlinecite{Dyak1},\onlinecite{Subashiev}). Though in both papers strain was assumed to be zero, this fact does not make a qualitative difference  when an original overlap of the light and heavy-hole bands is overcome by the size quantization. The only difference is that 
the energy due to the quantization compensates the negative strain energy such that the difference between the minima of the $lh1$ and $hh1$ sub-bands is small close to the transition point where $d=d_{|\Delta |}$. 
 Such a situation is assumed in Fig. \ref{size4}, where  
several  lowest sub-bands (closest to zero energy) are presented.  Once again, due to the inversion, the $hh1$ branch increases with energy instead of decreasing and at large enough $k_y $ transforms into the DK1 surface state. 
\par
We claim that Fig. \ref{size4} corresponds to the situation presented in Fig. 2(c) in  Ref. \onlinecite{Vail}. (We note that Fig. 2(b) of Ref. \onlinecite{Vail} demonstrates indeed that the width considered is already large enough, so that both branches of the Volkov-Pankratov states lie well below $\Gamma_8$ point).  The authors of Ref. \onlinecite{Vail} named the distance between the bottoms of the $lh1$ and $hh3$ sub-bands as "the bulk state band gap"  and the distance between the bottoms of the $hh1$ (DK1) and $hh2$ sub-bands as "the surface state band gap". We find that these quantities actually have different physical meaning.  We would say that the former quantity does not have any special physical meaning, but the latter should rather be named as "the bulk gap" of this quasi 3D system, see also Ref. \onlinecite{Dyak1}. 

\par
 Assume now that the conditions of Eq. (\ref{width}) are fulfilled, and, in addition, the Fermi level is within the $hh1$ (DK1) state, see  Fig. \ref{size4}.   Then despite the fact that the Fermi level lies below the bottom of the first light-hole sub-band ($lh1$), this DK1 state cannot be treated as a truly topological one since there is an appreciable overlap of the surface states wave functions located mostly at opposite boundaries (with the same spin orientations and opposite $k_y$). The reason for that is the fact that while keeping the Fermi level location below the bottom of the $lh1$ sub-band one cannot fulfill the condition $k_y d \gg \pi$ (see the discussion in the previous subsection). As a result, an elastic impurity scattering which keeps the spin orientation but changes the sign of $k_y$ will destroy the robustness of the surface state. 
\par 
The case of a relatively large negative strain when $d > d_{|\Delta |}$ and one has the DSM regime is presented in Fig.\ref{J} by the green curve (for a HgTe quantum well). At the transition point $d=d_{|\Delta |}$ the parameter $\Delta/\delta_h \approx 16$. Thus the green curve corresponds to $d \approx \sqrt{8} d_{|\Delta |}$. 
 If one takes $d$ to be much closer to $d_{|\Delta |}$, then the green curve will also be quite close to the blue one. 
Surely there is  again a strong overlap of the DK1 surface states wave functions located mostly at opposite boundaries. 

\section{Discussions}
\label{discus}
Next we discuss our results in the context of the work in Refs.~\onlinecite{Ohtsubo,Coster,Anh,Vail},
where different transitions with varying $d$ were studied both experimentally and theoretically.

For ultrathin Sn(001) films grown on InSb(001) substrates,
the authors of Ref. \onlinecite{Ohtsubo} propose the following scenario of surface-state band evolution observed in their ARPES data with increasing the film thickness $d$: 
(i) In the thin-film limit ($d\approx 2\,\text{nm}$), a gap is present in the band spectrum due to the hybridization between surface states localized at top and bottom surfaces. 
(ii) With increasing $d$, the gap closes and a Dirac cone appears in a small range of $d\approx (\text{3--4})\,\text{nm}$. The appearance of the Dirac cone is attributed to the decrease of the hybridization and formation of well-localized surface states on the top and bottom surfaces, signifying the topological regime. 
(iii) A further increase of $d$ to $d\approx 5\,\text{nm}$ leads to subbands of the heavy-hole character emerging near the Dirac cone and burying the latter in the continuum of the heavy-hole spectrum. The upper part of the Dirac cone remains intact, while the lower part hybridizes strongly with the heavy-hole subbands. 
(iv) At larger $d$, the bulk limit occurs, in which the heavy-hole and the light-hole bands touch each other forming a 3D zero-gap semimetal.
\par
The proposed scenario of surface-state band evolution differs from our findings obtained within the framework of the Kane model. We find that the VP states for $\alpha$-Sn have an extension $1/\kappa > 3.6\,\text{nm}$. 
The extension of $3.6\,\text{nm}$ is obtained in the case when the Dirac point lies in the middle of the band gap $E_g$, for which the VP states are maximally localized. The VP states can hardly be considered as non-overlapping even for the case (iii) with $d=5\,\text{nm}$. 
In other words, the overlap integral $J$ for these states is on the order of unity. 
For the case (ii), we find that the gap in the VP states is on the order of the fundamental bandgap $E_g $, which signifies absence of localization at the interfaces even for rather large values of the in-plane momentum ($k_y\sim E_g/P$). 
The physical situation described in case (ii) does not occur in the Kane model, because the heavy holes have a sufficiently large mass and do not undergo such a strong size quantization as to leave the VP states $S1$ and $S2 $ alone in the fundamental gap. 
For example, for $ d=3.5\,\text{nm}$, the heavy-hole subband $hh1$ shifts downwards in energy by $\approx 160$ meV, whereas $S2$ moves downwards by about the same amount. 
We find that the surface states are best pronounced in case (iv), where the role of the VP states is played by the DK states.
These states form due to the interaction between the VP states and the heavy holes, see Ref.~\onlinecite{Khaetskii}. 
The DK states have all the properties of topological states: absence of backscattering, spin-momentum locking, and single degeneracy.
\par
One could envision a scenario in which a large built-in electric field forms in the ultrathin $\alpha$-Sn film due to inequivalence of surfaces and associated charge transfers. 
In that case, the description of the splitting of the VP states derived in Sec.~\ref{VP-Gap} is modified. 
In essence, if the bias due to the electric field exceeds the splitting $\Delta_{\pm}$ given in Eq.~(\ref{gap1}), 
then two Dirac cones form, one per surface, with Dirac points shifted with respect to each other by the value of the bias. 
The splitting between the VP states on opposite surfaces occurs then at a finite in-plane momentum. 
The overall spectrum resembles two Rashba spectra, one dispersing upwards in energy and one downwards. 
Each of these Rashba-like spectra has a Dirac cone at zero in-plane momentum.
If the overlap integral is sufficiently small, the states detected on one of the two surfaces may appear as a pristine Dirac cone, 
because the states on the opposite surface are not visible in ARPES. 
Such an off-resonance scenario of interaction between the VP states on opposite surfaces 
might explain the unusual robustness of the Dirac cone observed in the experiment of Ref. \onlinecite{Ohtsubo}.
\par
In Ref.~\onlinecite{Anh},
epitaxial growth of $\alpha$-Sn on InSb with high quality is presented. 
The studied samples exhibit unprecedentedly high quantum mobilities of the surface states
(30 000 $\text{cm}^2/\text{V}s$), which is ten times higher than the previously reported values. 
The authors study the interfacial and bulk band structure of $\alpha$-Sn via 
the Shubnikov–de Haas oscillations combined with the 
first-principles calculations and claim that the results identify that  $\alpha$-Sn grown on
InSb is a topological Dirac semimetal (TDS).  Furthermore, a crossover
from the TDS to a 2D topological insulator and a subsequent phase transition
to a trivial insulator when varying the thickness of $\alpha$-Sn are demonstrated.
\par
We remark that the DSM--quantized 3D TI  transition,  which is one of the main interests of the current work, 
is not mentioned in Ref.~\onlinecite{Anh}. 
From the \emph{ab initio} modelling of the  $\alpha$-Sn/InSb structure, 
a very strong compressive in-plane strain for the $\alpha$-Sn layer is obtained in Ref.~\onlinecite{Anh}.
The strain value is five times larger than the value used by other research teams dealing with this system.
Nonetheless, the strain does not seem to affect significantly the sub-band energy separation,
or at least in ways that are expected from the Kane model.
Thus, in Fig. S8 of the Supporting Information of Ref.~\onlinecite{Anh}, 
the sub-band structure for the 56 ML-thick $\alpha$-Sn films under different in-plane strain 
(from strongly tensile to strongly compressive ones) is presented. 
The energy separation between the bottoms of the two lowest sub-bands with positive energies 
(which in our classification are $hh1$ and $lh1$) stays practically constant as a function of strain. 
From the Kane model, we expect that energy separation to  decrease to half of its value 
in going from Fig. S8 (a) to Fig. S8 (c). 
Recall that the gross effect of the strain is to shift the light-hole spectrum with respect to the heavy-hole spectrum,
since the two sectors separate from each other at zero in-plane momentum.
It is relatively easy to distinguish between a sparse ladder of light-hole sub-bands and a dense ladder of heavy hole sub-bands
and assess this gross effect of the strain in produced sub-band structure.
The strain effect obtained in Ref.~\onlinecite{Anh} seem to contradict this basic expectation.

\par
Besides, Ref.~\onlinecite{Anh} suffers from the same lack of correct level classification as Ref.~\onlinecite{Coster,Vail},
making the subsequent analysis confusing. 
For example, the authors believe that, 
due to quantum confinement, 
the bulk Dirac cone is gapped by $30\, \text{meV}$, see their Fig. 1b. 
In reality, this quantity is simply the energy separation between the first and second heavy-hole sub-bands
($hh1$ and $hh2$ in our labeling). 
Recall that the first heavy-hole sub-band $hh1$ in a quantum well is dispersing 
upwards in energy~\cite{Dyak1}.
The authors relate the upward-dispersing heavy-hole sub-band to the group of light-hole sub-bands (iLH in their notations).
As a consequence, the authors attribute the $30\, \text{meV}$ gap forming between two heavy-hole sub-bands to the gap 
of the bulk Dirac cone.
At the same time, the gap between the two heavy-hole sub-bands is present already at zero strain in the quantum well~\cite{Dyak1},
and therefore, that gap cannot be related to the bulk Dirac cone, because the latter form at a sufficiently strong negative strain.
We believe that the regime considered in Fig. 1b of Ref.~\onlinecite{Anh} corresponds to a sufficiently strong size quantization
between the light-hole and heavy-hole groups of sub-bands, such that the strain is insufficient to create a bulk Dirac cone, not even a precursor if it.
The necessary condition for the bulk Dirac cone to set in is $0 >E_{\textsl{lh1}}-E_{\textsl{hh1}}$, 
where $E_{\textsl{lh1}}$ and $E_{\textsl{hh1}}$ are the energies the first light-hole and the first heavy-hole sub-bands at zero in-plane momentum.
We stress again that both $E_{\textsl{lh1}}$ and $E_{\textsl{hh1}}$ disperse upwards in energy with in-plane momentum.
The energy $E_{\textsl{lh1}}-E_{\textsl{hh1}}$ is substantially larger than $30\, \text{meV}$ by about a factor of $6$.
\par
In summary, our results and conclusions differ significantly from those presented in Refs.~\onlinecite{Ohtsubo,Coster,Anh,Vail}.
These differences concern the correct interpretation of the sub-band structure in the quantum well 
as well as the identifications of 
all regimes in the presence of strain and quantization, and the transitions between them.
In particular, we assessed the robustness of the topological surface states in the region of the crossover between the DSM and quantized 3D TI phases, see Fig. \ref{J}, which was not done before.

\section{Conclusions} 
\label{conclude}
We have considered several problems in typical experimental situations related to topological surface states when a sample of the inverted-band material has a finite size and is subjected to strain.
The main interest is to understand the physics of the  series of transitions which happens at a given strain value with changing a sample width.  In particular, we have considered analytically the DSM -3D quantized topological insulator transition. It is believed to occur in the case when for a large  sample width one has a compressive in-plane strain which leads to an overlap of the light and heavy-hole bands (DSM regime).  We have identified  all the bulk-like and surface-like subbands which are involved in the transition, and, in particular,  
have shown that the branch of the surface states which is involved is the DK1 states (in strong contrast to the classification of the states used by the other researchers \cite{Ohtsubo,Coster,Anh,Vail}).  We have made concrete predictions about the robustness of the topological states near the transition point.   This conclusion opposes those reached by
the majority of the researchers. We have shown that near the transition point the surface state (DK1)
cannot be treated as a truly topological one. 
At least this state cannot exhibit the topological properties  in any transport experiments. 
The parameters of the problem near this point are such that an appreciable overlap of the surface states' wave functions located at opposite boundaries occurs. 
As a result, an elastic (and spin-conserving) impurity scattering between the states located at opposite boundaries will destroy the robustness of the surface state. 

\par 
We have also proposed a possible qualitative explanation of the discrepancy between the existing experimental ARPES data and  most theoretical predictions regarding the topological surface states spectra which is based on the pattern of a bending of the energy bands of the topological material near the interface  with the direct-band material. 

\par
 While solving these problems we have obtained several other interesting and important results. For example, we have derived the  effective boundary conditions for the solutions of the Kane model (with the contributions from the far bands) in the case of high barriers. It allows the use of this model to solve analytically various finite-size problems in the presence of strain for a realistic situation when heavy holes have finite masses. 
We hope that the results obtained can help in interpretation of current and future experimental data.

\par

This work is supported by the Air Force Office of Scientific Research (FA9550-AFOSR-23RYCOR05). 
 This research was performed while A. Khaetskii held an NRC Senior Research Associateship award at the Sensors Directorate, Air Force Research Laboratory. V.N.G. acknowledges financial support from the Spanish MCIN/AEI/10.13039/501100011033 
through the projects PID2020-114252GB-I00 (SPIRIT) and TED2021-130292B-C42, as well as from the Basque Government through the grant IT-1591-22 and the IKUR strategy program.
We thank S. Zollner for indicating to us Ref. \onlinecite{Cardona}. 

\appendix 
\section{Boundary conditions} 
\label{BCond}
\subsection{High-energy barriers between inverted and direct materials}
The wave functions (spinors) in all regions have the form 
\begin{eqnarray}
\psi = e^{ik_y y}\left( 
\begin{array}{ll}
u(x) & \\
v(x) & \\
w(x)
\end{array}
\right) 
\label{function}
\end{eqnarray}
Here $u$ corresponds to the electron contribution, while $v$ and $w$ describe the light-hole and heavy-hole contributions, respectively. We consider everywhere (except section \ref{DP_loc}) the case of high-energy offsets in the conduction and valence bands when the energy gap of the direct barrier material is large compared to the energy gap of the inverted-band material. Using this condition we find the following decaying solutions in the barrier region, $x>0$, for the heavy and light particles:
\begin{eqnarray}
\Psi_h=e^{-\kappa_{hb} x}\left( 
\begin{array}{ll}
0 & \\
0& \\
1
\end{array}
\right), \Psi_l=e^{-\kappa_{lb} x}\left( 
\begin{array}{cc}
-i \sqrt{\frac{1-\eta}{\eta}} & \\
1& \\
0
\end{array}
\right)
\label{barrierfunct}
\end{eqnarray}
Function Eq.(\ref{function}) in the barrier region is a linear superposition of the functions Eq.(\ref{barrierfunct}). 
Here $\kappa_{lb}=\sqrt{3\eta (1-\eta)/2}E_{gb}/P$, and $\kappa_{hb}=\sqrt{(1-\eta)E_{gb}/(\gamma-2\tilde\gamma)_b}$. 
The parameter $\eta$ ($0<\eta <1$) describes the asymmetry of the offsets in the conduction and valence bands, see Fig. \ref{Kiefer1}.  The edges of the conduction and valence bands in the barrier region are $E_{cb}=\eta E_{gb}$  and  $ E_{vb}=(\eta -1)E_{gb}$, where $E_{gb}>0$ is the energy gap of the direct-gap (barrier) material. The zero of the energy coincides with the middle of the 
$\Gamma_8-\Gamma_6$  gap of the inverted material. The case $\eta=1/2$ corresponds to symmetric offsets.  

\subsection{Effective boundary conditions}
\label{effect}
We start with the derivation of the boundary conditions (BCs) applicable for the case of arbitrary offsets in the conduction and valence bands. We use the standard procedure, i.e. we integrate all components of the Schrödinger equation with the Hamiltonian Eq. (\ref{Hamilt}) across the boundary $x=0$ from region 1 to region 2. The terms containing the $\gamma$ and $\tilde\gamma $ 
 we symmetrize in the standard manner, see Eq. (20) in Ch. III, Sec. II of Ref. \onlinecite{Bastard}. For example, the term 
$-2i\sqrt{3}k_y \tilde \gamma \hat{k}_x w(x)$ is presented as $-i\sqrt{3}k_y (\tilde \gamma \hat{k}_x w(x)+ \hat{k}_x (\tilde \gamma w(x)))$, etc. Then after intergation of the second and third equations we obtain
\begin{eqnarray}
\frac{-2iP}{\sqrt{6}}(u_2-u_1)+\gamma_{lb}v_2^{\prime}-\gamma_l v_1^{\prime}&=&\sqrt{3}k_y \delta \tilde \gamma w_1,  \nonumber \\
\sqrt{3}k_y \delta \tilde \gamma v_1 + \gamma_{hb} w_2^{\prime}-\gamma_h w_1^{\prime}=0, 
\label{bc0}
\end{eqnarray}
where $\delta \tilde \gamma =\tilde \gamma_b -\tilde \gamma$,  $\gamma_{hb}=(\gamma-2\tilde\gamma)_b$, etc., and we have already used that $v_1=v_2$, $w_1=w_2$.  The prime means the first derivative and indexes $1, 2$ refer to the well region 1 and the barrier region 2. From the explicit form of the barrier functions Eq. (\ref{barrierfunct}), we have the following equalities at $x=0$:  $u_2=-iv_1\sqrt{(1-\eta)/\eta}$, $v_2^{\prime}=-\kappa_{lb} v_1$, $w_2^{\prime}=-\kappa_{hb} w_1$, where again we used the continuity   of $v,w$ components. 
Then we obtain the {\it effective} BCs which contain only the quantities of region 1
 \begin{eqnarray}
\frac{-2iP}{\sqrt{6}}(-i\sqrt{\frac{1-\eta}{\eta}}v_1-u_1)-\gamma_{lb}\kappa_{lb} v_1 -\gamma_l v_1^{\prime}&=&\sqrt{3}k_y \delta \tilde \gamma w_1,  \nonumber \\
\sqrt{3}k_y \delta \tilde \gamma v_1 -\gamma_{hb}\kappa_{hb} w_1       -\gamma_h w_1^{\prime}=0.
\label{bc2}
\end{eqnarray}
We assume the condition $\gamma, \tilde\gamma \ll P^2/E_g$, where $E_g>0$ is the $\Gamma_8-\Gamma_6$ energy gap in the inverted-band material,  and similar condition in the barrier region.  
The functions $v_1,w_1$ in the region 1 have the form 
\begin{eqnarray}
v_1(x)=Ae^{\kappa_l x}a_2 +Be^{\kappa_h x}b_2, \nonumber \\
w_1(x)=Ae^{\kappa_l x}a_3 +Be^{\kappa_h x}b_3
\label{bc3}
\end{eqnarray}
Assuming in addition the high barrier conditions $k_y, \kappa_l, \kappa_h \ll \kappa_{lb}, \kappa_{hb}$, 
 one obtains the following {\it hard-wall effective} boundary conditions for the wave function inside the gapless material 1, see Fig. \ref{Kiefer1}:
\begin{equation}
\left (u+iv\sqrt{\frac{1-\eta}{\eta}} \right )|_{x=0}=0; \,\,\, w|_{x=0}=0 
\label{BCs}
\end{equation}
We note that at the opposite boundary ($x=-d$) the sign in front of the imaginary unity in Eq. (\ref{BCs}) should be changed for the negative one. 
\par
In real structures the values of $\gamma$ (or $\tilde \gamma$) parameters are of the same order of magnitude in both regions 1 and barrier region 2, i.e. $\gamma_l \simeq \gamma_{lb}$, for example. That is why the condition of the applicability of Eq. (\ref{BCs}) is 
\begin{equation}
E_g \ll E_{gb}\ll P^2 /\gamma
\label{applicab}
\end{equation}
\par
By  the effective boundary conditions we mean the following. We can consider the solution of the 
Schrodinger equation only in the inner region 1, all the information about the barrier regions enters the problem through the BCs  presented by Eq. (\ref{BCs}). Moreover, in the high barrier limit we avoid the usual troubles  related to the choice of the order of different noncommuting operators in the Hamiltonian,  Eq. (\ref{Hamilt}).  For example, in the popular book by G. Bastard \cite{Bastard} the author uses the very well-known BCs (a wave function and the current flux continuities) for the case of the finite offsets values. In this case one needs to take care about the order of the non-commuting operators in the Hamiltonian. G. Bastard uses the anti-commutator of the operators to satisfy the hermiticity of the Hamiltonian. However, it is the well-known fact that this combination is not unique, see also Ref. \onlinecite{Burt}. 
We should mention also that we are not aware of any work where the boundary conditions Eq. (\ref{BCs}) were obtained. 

\par   
It is remarkable  that even in the limit of very high barriers the BCs do not reduce to the trivial zero form ($u=0, v=0, w=0$), but contain the information about the asymmetry of the offsets in the conduction and valence bands. It is also important to mention that while considering the surface states problem with high-energy barriers and for energies near the $\Gamma_8$ point (i.e. much smaller than the energy gap $E_g$) 
one can neglect the contribution from the electron $\Gamma_6$ band.
Then the boundary conditions within the Luttinger model reduce to $v=0,w=0$, i.e. to zero values of the light-hole and heavy-hole components \cite{Dyak,Dyak1}. 
\par 
To assess how realistic
the approximation of the high barrier is for the real material CdTe/HgTe, we can mention the following. The general condition of the high barrier requires the following inequalities, 
$\eta E_{gb} \gg E_g, (1-\eta) E_{gb} \gg E_g$, 
see Fig. \ref{Kiefer1}. It means that the decaying factor $\kappa$ of the wave function of the light particles under the barrier should be much larger than in the well region (HgTe). (For the heavy holes this factor is much larger than for the light particles). This quantity $\kappa$ in the material region has totally different value for the VP   surface states and for the DK states.  
\par
  In the case of the VP states we can estimate the location of the Dirac point of the VP states, Eq. (\ref{DP}), in two cases: for the infinite barrier and for the real parameters of the CdTe/HgTe structure and zero strain. 
 For the real structure one has $\epsilon_{DP}$=61 meV  compared to $\epsilon_{DP}$=72 meV in the infinite barrier case. We have used $\eta$ =0.74. Thus it makes an error of less than 4 {\%} for the position of the Dirac point on the scale of $ E_g$. In absolute units, the error amounts to an energy shift of about 10 meV, for surface states buried by $\sim $100 meV  under the closest bulk band edge.
\par
  As for the DK states, which play the major role in the descriptions of physics in \ref{Wing} and \ref{Interplay} sections, the situation is much better. The reason is that the energy of the DK1 state in the HgTe region is close to the edge of the $\Gamma_8$- band, and the corresponding factor $\kappa$ for that region can be arbitrary small   (depending on the $k_y$ value). 

\section{Correction to Eq. (\ref{DP}) due to finite values of $\gamma $ and $\tilde\gamma $}
\label{D1}
Let us derive the correction to Eq. (\ref{DP}) due to interaction with the far bands, i.e. we will take into account the terms proportional to $\gamma$ and $\tilde \gamma$  in the Hamiltonian,  Eq. (\ref{Hamilt}). For $k_y=0$ the heavy holes are completely decoupled from the light particles, and to find the location of the DP of the Volkov-Pankratov states we need to solve the 2x2 problem (the top left corner of the matrix Eq. (\ref{Hamilt})). 
The wave function is a two-component spinor
\begin{eqnarray}
\psi = \left( 
\begin{array}{ll}
u(x) & \\
v(x) 
\end{array}
\right) 
\label{spinor1}
\end{eqnarray}
To obtain the boundary conditions (BCs) we use the standard procedure, i.e. we integrate the both components of the Schrödinger equation across the boundary $x=0$. The term $\gamma_l \hat{k}_x^2$ ($\gamma_l=\gamma+2\tilde\gamma $)
 we symmetrize in the standard manner: \cite{Bastard} $-\partial/\partial x \gamma_l \partial/\partial x $. Then we obtain the following BCs
\begin{eqnarray}
\frac{-2iP}{\sqrt{6}}(u_2-u_1)+\gamma_l^{(2)}v_2^{\prime}-\gamma_l^{(1)}v_1^{\prime}=0, & 
\nonumber \\
v_1=v_2,  
\label{bc1}
\end{eqnarray}
where the prime means the first derivative and indexes $1, 2$ refer to the regions $1,2$. The wave functions decay in the left region 1 as $\exp (\kappa_1 x)$ and in the right region 2 as $\exp (-\kappa_2 x)$, with positive $\kappa_{1,2} $. The values of these quantities are (for $\Delta=0$)
\begin{equation}
\kappa_i^2=\frac{(\epsilon_{ci}-\epsilon)(\epsilon-\epsilon_{vi})}{2P^2/3 + \gamma_l^{(i)}(\epsilon_{ci}-\epsilon)}; i=1,2 
\label{kap}
\end{equation}
Using the BCs Eq. (\ref{bc1})  we finally obtain 
\begin{equation}
\frac{2}{3}P^2 \left[ \kappa_1 +\kappa_2 \frac{(\epsilon-\epsilon_{c1})}{(\epsilon-\epsilon_{c2})} \right ]=(\epsilon-\epsilon_{c1})(\kappa_1 \gamma_l^{(1)}+\kappa_2 \gamma_l^{(2)})
\label{answer1}
\end{equation}
Eq. (\ref{kap}) and Eq. (\ref{answer1}) should be solved together to find the wanted energy $\epsilon$.
In the first order with respect to small quantities $\gamma_l^{(1,2)}$, writing $\epsilon=\epsilon_0 +\delta \epsilon$, one obtains 
\begin{equation}
\delta \epsilon =\frac{1}{\left[1-\frac{\epsilon_0 (\epsilon_0-\epsilon_{c2})}{\kappa_{10}\kappa_{20}2P^2/3}\right ]}
\frac{(\kappa_{10} \gamma_l^{(1)}+\kappa_{20} \gamma_l^{(2)})}{(\kappa_{10}+\kappa_{20})4P^2/3}
(\epsilon_0-\epsilon_{c1})(\epsilon_0-\epsilon_{c2})
\label{first_order}
\end{equation}
Here 
$$
\kappa_{10}^2=\frac{(\epsilon_{c1}-\epsilon_0)(\epsilon_0-\epsilon_{v1})}{2P^2/3}, \, 
\kappa_{20}^2=\frac{(\epsilon_{c2}-\epsilon_0)(\epsilon_0-\epsilon_{v2})}{2P^2/3},
$$
and $\epsilon_0$ is given by Eq. (\ref{DP}) for $\Delta=0$. Shift $\delta \epsilon $ is negative for both cases we consider in this work (HgTe/CdTe and $\alpha $-Sn/InSb). 
To illustrate Eq. (\ref{first_order}) we present here the answer for the case of the structure with the symmetric offsets ($\eta=1/2$) when $\epsilon_0=0$
\begin{equation}
\delta \epsilon =-
\frac{3(E_g \gamma_l^{(1)}+E_{gb} \gamma_l^{(2)})}{16P^2}\frac{E_g E_{gb}}{E_g+E_{gb}}
\label{eta_1/2}
\end{equation}
Note that here the following parameters are assumed to be small: $\gamma_l^{(1)}E_g /P^2 \ll 1$,  $\gamma_l^{(2)}E_{gb}/ P^2 \ll 1$.

\section{Derivation of   Eq. (\ref{gap1})} 
\label{A1}

We derive here Eq. (\ref{gap1}) and  also obtain a shift of the central point of the VP branches due to finite size effects in the limit of large width ($\alpha/s \gg 1$). 
We rewrite Eq. (\ref{gap}) in the following form 
\begin{eqnarray}
(1+2\epsilon \rho/E_g) (1-2e^{-2\alpha \xi})=\xi /s,  \nonumber \\
\xi(\epsilon)=\sqrt{1-4(\epsilon/E_g)^2+2P^2k_y^2/3E_g^2}, 
\label{start}
\end{eqnarray}
and start with the case $k_y=0$. Introducing $x=2\epsilon/E_g $, we look for the small correction $\delta x $, 
$x=-\rho +\delta x$, where $x_{\infty}=-\rho $ and $\xi_{\infty}=1/s $ is the solution of the zeroth order problem ($d=\infty $), see Eq. (\ref{VP_infty}). Assuming that the leading order of $\delta x$ is proportional to $ e^{-\alpha /s}$, and keeping in the left hand side of Eq. (\ref{start}) only the terms which contain the power of the exponent not higher than three 
($e^{-3\alpha /s})$, we obtain
$$
(1-2e^{-2\alpha/s})(1+\rho s^2 \delta x) +4\alpha s\rho e^{-2\alpha/s} \delta x=\sqrt{1+2\rho s^2 \delta x -s^2 \delta x^2}
$$
By squaring both sides of this equation (in this way we only need to keep the terms proportional to $\delta x $ and $\delta x^2$) we finally easily obtain
\begin{equation}
\delta x^2 -\frac{8\rho}{s^2}(1-\frac{\alpha}{s}) e^{-2\alpha /s} \delta x  -\frac{4}{s^4}e^{-2\alpha /s} =0
\label{correction}
\end{equation}
The first and the last terms determine the gap value, and the second  one gives the shift of the central point of the  VP branches 
\begin{equation} 
\frac{\delta \epsilon}{E_g} = \pm \frac{1}{s^2}e^{-\alpha/s} + \frac{2\rho}{s^2}e^{-2\alpha/s}(1-\frac{\alpha}{s})
\label{shift}
\end{equation}
Actually, under the condition we  have used while deriving Eq. (\ref{shift}), the term in the brackets proportional to $\alpha$ prevails. The shift is proportional to $e^{-2\alpha/s}$ and  is parametrically smaller than the gap value. 
\par
While deriving Eq. (\ref{gap1}) we can neglect the shift value, as a result we can take the exponential function in the left hand side of Eq. (\ref{start}) in the form $\exp(-2\alpha \xi_{\infty})$ ($\xi_{\infty}$ is given by  Eq. (\ref{VP_infty})). Then the problem reduces  to a solution of the quadratic equation for $\epsilon$, which gives the result Eq. (\ref{gap1}), where we have neglected  the exponentially small quantities everywhere except in the expression for $\Delta_{\pm}$.

\section{Exact equation for $k_y^{\star}$ obtained within the Luttinger model for a  finite-size sample in the presence of  strain}
\label{A2}

We consider a film of the finite width $d$ and use zero boundary conditions on both surfaces. 
The exact equation that describes the critical value of $k_y=k_y^{\star}$ where the quantized bulk-like $hh1$ state  transforms into the surface DK1 branch reads
\widetext
\begin{eqnarray}
\frac{\tanh(\kappa_h d/2)}{\kappa_h d/2}=\frac{\{ 4\tilde\gamma^2k_y^2[\epsilon+\Delta/4-(\gamma-\tilde\gamma)k_y^2]^2-\kappa_h^2[\tilde\gamma \epsilon +\gamma\Delta/4 -
(\gamma-\tilde\gamma)2\tilde\gamma k_y^2]^2  \}}{4\tilde\gamma^2k_y^2 [\epsilon+\Delta/4-(\gamma-\tilde\gamma)k_y^2][\epsilon+\Delta/4-(\gamma-\tilde\gamma)(k_y^2-\kappa_h^2)]} 
\nonumber \\
\kappa_h^2=2k_y^2 + \frac{2(\gamma \epsilon +\tilde\gamma \Delta)}{(4\tilde\gamma^2-\gamma^2)}
\nonumber \\
\epsilon=\epsilon_{HH}=\gamma k_y^2 -\sqrt{\Delta^2/4 +4\tilde\gamma^2k_y^4-\tilde\gamma k_y^2 \Delta}
\label{kstar}
\end{eqnarray}
\endwidetext
Note that in this Appendix the definition of $\gamma, \tilde \gamma$ is different from the one in the rest of the paper. Here these constants are "full" Luttinger parameters, thus $2\tilde \gamma -\gamma >0, 2\tilde \gamma +\gamma >0$. Eq. (\ref{kstar})
is obtained from the condition $\kappa_l=0$.  The expression for $\kappa_h$ presented above follows  from the equation 
\begin{equation}
k_y^2-\kappa^2_{l,h}=\frac{\gamma \epsilon +\tilde{\gamma}\Delta \mp \sqrt{(\gamma \Delta/2+2 \tilde{\gamma}\epsilon)^2+ 3\tilde{\gamma}\Delta k_y^2 (4\tilde{\gamma}^2-\gamma^2)}}{(\gamma^2-4\tilde{\gamma}^2)}
\label{kappa}
\end{equation}
using $\kappa_l=0$ as identity. 
Note that Eq. (\ref{kappa}) coincides with Eq. 22 of Ref. \onlinecite{Khaetskii} and 
the quantity $\epsilon_{HH}$ in Eq. (\ref{kstar}) is nothing but the edge of the bulk heavy-hole band in the presence of strain. 
 Therefore, at the point $k_y^{\star}$ the  heavy-hole components of the $hh1$ state transform from the form $\sin k_x x, \cos k_x x$ (localized near the center of the film) to the exponents which are localized more near the boundaries (DK1 state).
 It is possible to show that there is only one solution of Eq. (\ref{kstar}) at $\Delta \geq 0$, and for $\Delta=0$ its solution $k_y^{\star}$ coincides with the one obtained from the equations of Ref. \onlinecite{Dyak1}. In the case $d \to \infty$ 
the corresponding solution tends to the value given by Eq. (27) of Ref. \onlinecite{Khaetskii}.


\begin{thebibliography}{99}
\bibitem{noteintegral}
Strictly speaking, the overlap integral $J$ is a 2x2 matrix, because of the spin degeneracy in the symmetric quantum well.
However, that matrix can be reduced to a scalar due to the time-reversal symmetry of the Hamiltonian 
after  a suitable choice of the basis.
\end{thebibliography}
\end{document}